\documentclass[fleqn,10pt]{wlscirep}
\usepackage[utf8]{inputenc}
\usepackage[T1]{fontenc}
\usepackage{lineno}
\usepackage[numbers]{natbib}
\usepackage{wrapfig}
\usepackage{booktabs}
\usepackage{mathtools}
\usepackage{nccmath} \usepackage{soul} \usepackage{float}

\title{Using AI to Summarize US Presidential Campaign TV Advertisement Videos, 1952--2012}

\author[1,*]{Adam Breuer}
\author[3,*]{Bryce J. Dietrich}
\author[2,6]{Michael H. Crespin}
\author[5,]{Matthew Butler}
\author[2,]{JA Pryse}
\author[4]{Kosuke Imai}

\affil[1]{Dartmouth College, Department of Computer Science and Department of Government, Hanover, NH, USA}
\affil[2]{University of Oklahoma, Carl Albert Congressional Research and Studies Center, Norman, OK, USA}
\affil[3]{Purdue University, Department of Political Science, West Lafayette, IN, USA}
\affil[4]{Harvard University, Department of Government and Department of Statistics, Cambridge, MA, USA}
\affil[5]{University of Iowa, Digital Scholarship \& Publishing Studio, Iowa City, IA, USA}
\affil[6]{University of Oklahoma, Department of Political Science, Norman, OK, USA}
\affil[*]{corresponding authors: Adam Breuer (adam.breuer@dartmouth.edu) and Bryce J. Dietrich (bjdietri@purdue.edu)}

\begin{abstract}
This paper introduces the largest and most comprehensive dataset of US presidential campaign television advertisements, available in digital format. The dataset also includes machine-searchable transcripts and high-quality summaries designed to facilitate a variety of academic research. To date, there has been great interest in collecting and analyzing US presidential campaign advertisements, but the need for manual procurement and annotation has led many to rely on smaller subsets. We design a large-scale, parallelized, AI-based analysis pipeline that automates the laborious process of preparing, transcribing, storyboarding, and summarizing videos. We then apply this methodology to the 9,707 presidential ads from the Julian P. Kanter Political Commercial Archive. We conduct extensive human evaluations to show that these transcripts and summaries match the quality of manually generated alternatives. We illustrate the value of this data by including an application that tracks the genesis and evolution of current focal issue areas over seven decades of presidential elections. Our analysis pipeline and codebase also show how to use LLM-based tools to obtain high-quality summaries for other video datasets.
\end{abstract}

\begin{document}

\flushbottom
\maketitle

\thispagestyle{empty}

\section*{Background \& Summary}

Studies of political campaign advertisements help us understand what messages political elites send to voters and how opposing campaigns strategically interact.  Campaign ads also play a critical role in identifying salient issues that matter to the electorate.

In the United States, television ads are the primary means through which campaigns reach voters. In the $2024$ US election alone, political campaigns aired ads on broadcast television more than $1.47$ million times at a cost of over $5.3$ billion dollars between Super Tuesday and November $1$st \cite{AdImpactPlay, AdImpactReview}. Despite the recent trend towards online advertising, broadcast television advertising continues to increase in real amounts year-over-year. Television also continues to dwarf alternatives, accounting for $48\%$ of ad spend, more than triple the value of online and social media advertising campaigns \cite{AdImpactReview, borrell2017}. Therefore, it is no surprise that TV advertisements for the US presidential campaign have been a focal point of considerable research in political communication \cite{jamieson1996packaging, kaid2004political}, political psychology \cite{brader2006campaigning, biocca2013television, valentino2004impact} and political behavior \cite{goldstein2002campaign,geer2008defense,gerber2011large, fowler2018political}.

Despite this scholarly interest, two challenges remain that hinder further development of studies on US presidential campaign TV ads.  First, the literature on ads has repeatedly emphasized the importance of analyzing how the content and style of ads have evolved over time~\cite[e.g.,][]{petrocik2003issue,hayes2005candidate}. However, such analysis is largely precluded by the fact that ad data are limited or unavailable for elections prior to 1996. Second, even when ads from recent elections are available, extracting relevant information from the videos for downstream analyses has required costly human hand-coding, limiting the scope of analyses to a specific set of variables 
(see the Wesleyan Media Project \citep{fowler2016political}
and its predecessor Wisconsin Advertising Project \citep{goldstein2002political, goldstein2007congressional}).  
In part due to these limitations, scholars have focused primarily on analyzing small subsets of US presidential campaign TV ads \cite[e.g.,][]{prior2001weighted}.

\subsection*{Overview of the dataset: Ad videos, transcripts, and summaries, 1952-2012}

We address these problems by publicly releasing the largest collection of US presidential campaign TV advertisements to date from the Julian P. Kanter Political Commercial Archive \cite{kanterpcc}. There are $9{,}707$ ads for presidential candidates starting with the $1952$ election where Dwight D. Eisenhower faced Adlai Stevenson and ending with the $2012$ Obama v. Romney election. Although the ads primarily feature major candidates, the data also include commercials from lesser-known candidates who did not garner the nomination, permitting analysis of campaign messaging on either side of primaries and major party nominations. 

To facilitate research on this dataset, we develop a large-scale, parallelized AI-based analysis pipeline that automates the laborious process of preparing, transcribing, and summarizing videos.  The proposed methodology is designed to facilitate the neutrality, comprehensiveness, and coherence that are desirable for downstream academic research. We then apply this analysis pipeline to the ad dataset and release each ad with an accompanying high-quality transcript and an efficient $50$-word summary. In addition to these transcripts and summaries, we also provide extensive metadata and an online search engine so that researchers can easily access the ads, determine the subsets of ads that are relevant to their research topic or time period, and use them in their downstream analyses.  

At the time of writing, LLMs that are capable of summarizing video files are not yet available.  Therefore, our analysis pipeline itself may be of independent interest, as it permits researchers to efficiently generate neutral and comprehensive summaries for each video in a large video dataset. The proposed methodology improves upon a recent study, which shows that automated coding using machine learning algorithms performs as well as human coding \cite{tarr:hwan:imai:23}.  Compared to that study, our transcripts are of higher quality. Moreover, instead of just coded variables, we are also able to provide a natural language summary of each ad's audio \textit{and} visual content that is designed to facilitate efficient data exploration.

We validate each step of our pipeline used to generate quality AI-generated summaries. Specifically, we validate our video preprocessing step, AI video transcription step, and AI video summarization step by comparing stratified samples of these outputs with human-generated outputs. Our main validation result is that human specialists find the quality of our LLM-generated ad transcripts and summaries to be at least equal to, and often superior to, the quality of transcripts and summaries composed by human research assistants across multiple key dimensions. Importantly, we also show that the high quality of our transcripts and summaries remains consistent across key sources of variation in the literature, including political party and time. This result and several others from the validation exercises outlined below will give applied researchers a great deal of confidence when they use both the raw ads and the associated summaries and transcripts for their own research.

Finally, we also substantively validate and illustrate our dataset by considering an application that measures the historical evolution of current key electoral issue areas, as viewed through a keyword-assisted topic model estimated on the transcripts of all ads in the dataset. We show that the seven decades of data exhibit trends in campaign issues that correspond to historically recognized developments in campaign issue agendas.

\subsection*{History of the Kanter campaign advertisement video collection}

A former staffer for Adlai Stevenson, Julian P. Kanter started the Kanter political TV ad collection in $1956$ when he realized that political commercials were being discarded, or later taped over, after airing. He obtained the early film ads from radio and television studios, ad agencies, and the campaigns themselves. He later traded blank tapes in exchange for the already aired commercials. The collection moved to the University of Oklahoma in $1986$ when it outbid the Smithsonian to purchase approximately 25,000 radio and television ads from Julian Kanter for \$250,000 \cite{harp201660}, including approximately 21,000 television ads ranging from 1950 to 1984 \cite{haynes1996political}. The archive has increased in size as Kanter and others continued collecting ads and political advertising firms donated their film and video. At the University of Oklahoma, the archive was originally housed in the Political Communications Center and is now part of the Carl Albert Congressional Research and Studies Center’s archival collections \cite{pryse2022practical}.
Unfortunately, there are no presidential ads available in the dataset beyond the 2012 election. This is largely due to a resource constraint within the archive itself that limits active collecting and the reluctance of political consultants to donate commercials for current or potential future active candidates. We hope to add these recent ads to the dataset in the future and process them by applying the same techniques described in this paper.  In the meantime, the missing ads limit what scholars are able to say about the last three elections in 2016, 2020, and 2024. 

\begin{table}[t]
\centering
\tabcolsep=3.4pt
\small
    \addtolength{\leftskip} {-2cm} \addtolength{\rightskip}{-2cm}
\begin{tabular}{@{}|l||r|@{}}
\toprule
Ads       &   $9{,}707$  \\ 
     \midrule
Elections       &   $1952$ - $2012$  \\ 
     \midrule
Democratic Ads       &    $5{,}859$  \\ 
     \midrule
Republican Ads       &    $3{,}848$  \\ 
     \midrule
Total Hours of Video       &   $102.95$  \\ 
     \midrule
Format       &   mp4: \ $720\times480$  \\ \bottomrule
\end{tabular}
    \caption{Summary of the Presidential Ad dataset.}
\label{tab:summarystats}
\end{table} 
\subsection*{The US presidential advertisement video dataset}

Although the collection has grown to more than $100{,}000$ commercials from campaigns up and down the ticket, we focus on the $9{,}707$ ads for the US presidency from $1952$-$2012$. Table~\ref{tab:summarystats} reports an overall summary of the ad collection.  We do not make any claim that the collection contains the \textit{complete} universe of ads (and it is not known which ads are missing), or that it is a representative sample of all ads.  However, it is the most complete collection available in such a long time frame. Indeed, for a substantial number of items in the collection, the archive has the only known copy anywhere. To emphasize this point, we asked two undergraduate research assistants to check whether the ads in our dataset also appear in 
\textit{The Living Room Candidate} \citep{clinton1992living}, 
which is the largest known collection of presidential ads (365 ads matching our $1952$-$2012$ time period) available online. 
Using a random sample of size $N=500$ from our collection, they found only $10.6$ percent of our ads were available on that website.

Over the years, the collection has been used or cited not only in political science research \cite{kaid1991negative, finkel1998spot, damore2003using, prior2007post, geer2008defense, brader2020campaigning}, but also in \textit{amicus curiae} briefs \cite{schofield2016amicus}, documentaries \cite{power, octopus}, podcasts \cite{wilder}, and in many academic disciplines, including advertising \cite{beard2020advertising}, area studies \cite{dover2002videostyle}, cinematography \cite{bernard2020archival}, communication \cite{biocca1991television, dalton2011third, biocca2013television, biocca2014television, dunaway2019effects}, history \cite{webb2008freedom, nelson1989sources}, and library and information science \cite{haynes1996political, albitz2001locating}. Nevertheless, access to the data has been significantly limited by the fact that until recently, anyone wishing to examine a substantial number of Kanter ads had to visit the archives in person. Our goal is to remove this barrier and make the dataset both widely available to scholars and also amenable to efficient research curation via the release of accompanying transcripts, summaries, and an online interface that permits keyword searches that return the corresponding relevant subset of ads.

\subsection*{Transcripts}
In addition to the ads themselves, we also release a high-quality transcript for each ad. These transcripts serve three purposes. First, they permit practitioners to search the ad dataset for keywords of interest in order to identify ads that are relevant to their specific research topic. We provide online text search functionality on our project website \url{https://campaignTVads.org/} (see \textit{Usage Notes} below). Second, transcripts allow practitioners to study the dataset using novel methods from the recent renaissance in text-based research methods for social science \cite[see e.g.,][]{grimmer2022text, eshima2024keyword, egami2022make, fong2023causal}. Finally, transcripts are also a key ingredient of our ad summarization workflow, as we describe below.

\subsection*{50-word summaries and visually representative `storyboard' still frames}  Historically, most research on political ads has focused on small subsets of ads, in no small part due to the high manual labor costs associated with viewing ads. Specifically, while each individual ad is relatively brief, our dataset contains more than $100$ hours of ads in total, which can prevent practitioners from conducting a thorough data exploration. Transcripts are helpful in this regard, but reading transcripts requires nearly as much time as watching videos and also ignores the rich visual information in the ads. 

To address this problem, we take inspiration from a rich scholarly tradition of `campaign ad storyboards'---time-synced transcript segments associated with a visually representative selection of still frames extracted from each campaign ad (see, e.g., \citep{freedman1999measuring, goldsteinstoryboardbook}). Specifically, to supplement the ad transcripts, we develop a novel method to automatically extract a set of still frames that comprehensively represent the visual contents of each ad. We include these `storyboard' still frames alongside each ad and transcript in our dataset.
We then use the transcripts and these visually representative still frames to generate a high-quality $50$-word summary of each ad designed to significantly accelerate the process of manually parsing the dataset. These summaries offer an efficient synopsis of \textit{both} audio and visual content, and each summary can be parsed by a human in approximately $4$ seconds.

\subsection*{Metadata}
Finally, we also provide additional metadata about each ad. Our files include the candidate name, political party, election, ad title (if known), original video technology format (e.g., 16mm film), and ad duration.  

\section*{Methods}
\label{sec:methods}

In this section, we discuss the proposed automated procedure that takes a historical campaign TV ad and produces a complete transcript, a visually representative selection of still frames, and an efficient, high-quality $50$-word AI-generated summary designed to facilitate downstream research.  We begin by describing how we preprocessed video files, as the original raw files contained extra segments at the beginning and end that were not part of the ad.  We then show how to generate ad summaries using LLMs. 

Importantly, our summarization pipeline serves as readily-adaptable model. By providing the pipeline and data for all intermediate outputs, such as transcripts and key frames, we enable other researchers to generate their own custom-directed summaries or ad labels merely by adjusting the provided prompts and recomputing the final step of the pipeline, bypassing the significant cost and complexity of preprocessing, transcription, and frame extraction steps and their associated human validation workflows.

We illustrate our approach throughout this section by considering the running example of Bill Clinton's famous 1992 `Hope, Arkansas' ad (ID P-$1291$-$61062$). This ad (\url{https://vimeo.com/992368244}) is a challenging example as it contains a diverse set of themes and images.

\subsection*{Preprocessing videos to extract political ads.}
\label{ssec:preprocessing}

Recordings of historical television ads typically contain `pre-roll' and/or `post-roll' clips that include, for example, the ad metadata, the name of those who recorded the ad, and the television station.  It is important to trim these segments so that they do not hinder efficient viewing, and so their content is not included in the downstream ad summary we generate. However, automatic detection of transition points between the pre/post-roll clips and an ad itself is a challenging task because these segments vary widely in terms of length, format, and content.

We develop an automated trimming algorithm based on the following key observation; despite their diversity, nearly all political ads begin and end with a scene that contains video footage of people and/or audio of speech, such as voiceover or a fading in (or out) scene depicting a candidate or interviewer speaking. 
We leverage this observation by performing, on each video, (1) automated video detection of the timestamps of spoken words, (2) automated video person detection, and (3) automated video scene detection, using the Google Video Intelligence API \cite{googleAPI}. As an aside, we note that researchers who wish to apply our methodology to their own datasets may skip using Google Video Intelligence API for spoken word timestamp detection and simply use the Whisper beginning/ending timestamps instead, as they are computed in \textbf{Step~1} below anyway.
We then trim the beginning and end segments of the video file, yielding the video clip that spans from the beginning of the earliest scene that contains a person or spoken word to the end of the last scene that contains a person or spoken word. 

\begin{figure}[t]
\centering
\includegraphics[width=1\textwidth]{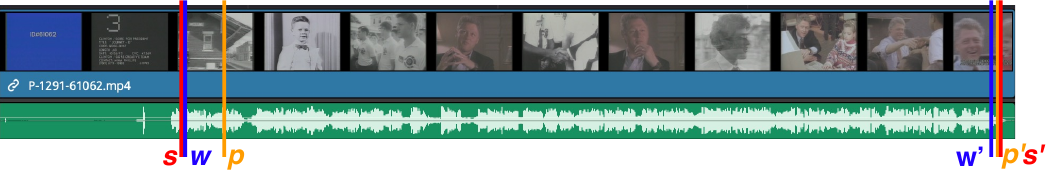} 
\caption[Preprocessing]{Preprocessing example: Time stamps $w$ and $w'$ are the start of the first spoken word and the end of the last word spoken, respectively. Points $p$ and $p'$ are the start of the first person and end of the last person depicted.  Lastly, $s$ and $s'$ are the start of the first detected scene and the end of the last detected scene that contain a word or person, respectively.  We keep the segment from $s$ to $s'$ as the advertisement, and we trim on either side.}
\label{fig:preprocessing}
\end{figure}

Fig.~\ref{fig:preprocessing} illustrates this process via an example.  Here, $w$ represents the start of the first spoken word, and $p$ indicates the moment the first detected person appears onscreen. We seek $s$, which is the beginning of the first scene that contains $w$ or $p$. Similarly, $w^\prime$ and $p^\prime$ represent the end of the last spoken word and the time the last detected person leaves the screen. Finally, $s^\prime$ is the endpoint of the last scene that contains a word or person, i.e., the last scene that contains $w^\prime$ or $p^\prime$.  In this case, we take $[s, s^\prime]$ as the ad segment, which contains all of these six points.

\subsection*{Generating ad summaries that limit mismeasurement-by-omission}
\label{ssec:summariesoverview}

Our video summarization workflow is designed to overcome two primary challenges. First, at the time of writing, mainstream LLM tools accept audio and image inputs, but they \textit{lack the capability to accept video inputs}. Second, even if video-processing LLM tools were available, our use case does not directly conform to popular video description tasks' objectives.  Specifically, because our video summaries may later be used for downstream statistical analysis, our primary objective is not to generate summaries that are articulate or interesting. Instead, our main goal is to ensure descriptive completeness and minimize potential sources of bias and mismeasurement.  In particular, we do not want to omit important descriptive information. 

\begin{figure}[t]
\centering
\includegraphics[width=0.45\textwidth]{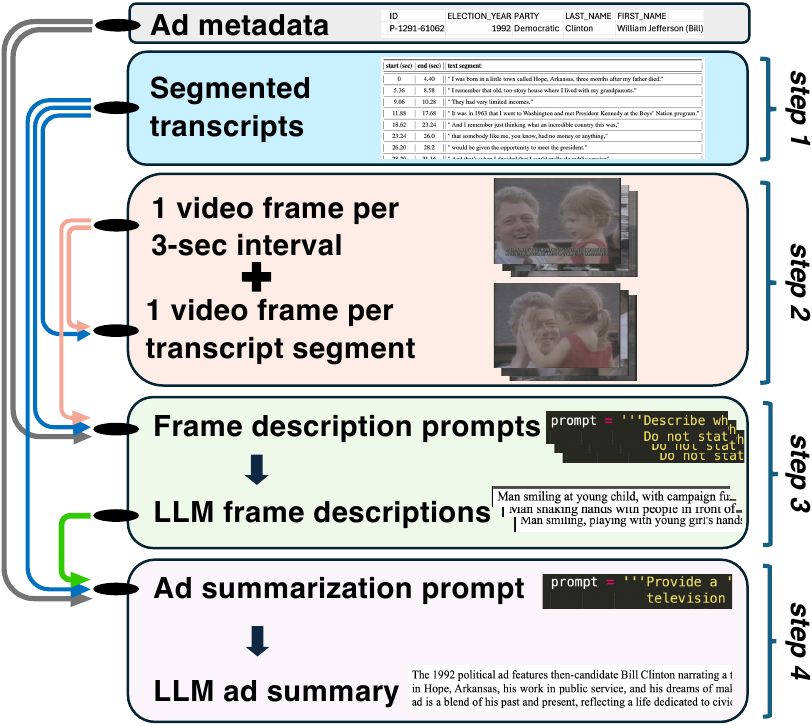} \caption[Workflow]{Ad video summarization workflow. Dependencies between intermediate outputs are indicated by arrows on the left.}
   \label{fig:workflow}
\end{figure}

We design a summarization workflow that both ($i$) accepts video inputs and ($ii$) limits bias and omission errors that might affect downstream analyses. As shown in Fig.~\ref{fig:workflow}, we address these two challenges by dividing the summarization of each video into three independent tasks designed to favor descriptive completeness at the expense of redundancy: (\textbf{Step~1}) transcription, (\textbf{Step~2}) video segmentation, and (\textbf{Step~3}) frame description. We then merge these many (possibly redundant) summaries into a single comprehensive ad summary (\textbf{Step~4}).  We now describe each step in detail.

\subsubsection*{Step 1: Transcription and transcript segmentation}

After video preprocessing, we perform automated transcription and transcript segmentation using the largest and most performant version of OpenAI's LLM-based Whisper transcription model: \textsc{whisper-large-v3} \cite{radford2023robust}. This model improves upon a previous version (\textsc{v2}) that already obtained transcription accuracy within a fraction of a percent of professional human transcribers across a variety of text benchmarks in terms of word error rates \cite{radford2023robust}. 

The Whisper model is open source, which permits us to parallel-process many videos simultaneously on the local resources of a cloud-based server computer. Specifically, a standard Google Cloud Compute virtual machine with 48 cores transcribes our approximately 10,000 videos in a couple of hours by running parallel batches of $48$ videos at a time. 

\begin{table}[t]
\centering
\tabcolsep=3.4pt
\small
    \addtolength{\leftskip} {-2cm} \addtolength{\rightskip}{-2cm}
\begin{tabular}{@{}|c|c||l|@{}}
\toprule
     \textbf{start (sec)} & \textbf{end (sec)}     & \textbf{text segment}:  \\ \midrule \midrule
0   &  4.40      & " I was born in a little town called Hope, Arkansas, three months after my father died." \\ \midrule
5.36   &  8.58      & " I remember that old, too-story (\emph{sic}) house where I lived with my grandparents."   \\ \midrule
9.06   &  10.28     & " They had very limited incomes." \\ \midrule
11.88   &  17.68      & " It was in 1963 that I went to Washington and met President Kennedy at the Boys' Nation program."  \\ \midrule
18.62   &  23.24      & " And I remember just thinking what an incredible country this was," \\ \midrule
23.24   &  26.0      & " that somebody like me, you know, had no money or anything,"  \\ \midrule
26.20   &  28.2      & " would be given the opportunity to meet the president."  \\ \midrule
28.20   &  	31.16      & " And that's when I decided that I could really do public service"   \\ \midrule
	31.16   &  	32.64      & " because I cared so much about people."   \\ \midrule
33.18   &  36.56      & " I worked my way through law school with part-time jobs, anything I could find."   \\ \midrule
37.4   &  	40.46      & " And after I graduated, I really didn't care about making a lot of money."   \\ \midrule
	40.54   &  42.68      & " I just wanted to go home and see if I could make a difference."   \\ \midrule
	43.72   &  48.22      & " We've worked hard in education and health care to create jobs,"   \\ \midrule
48.36   &  50.16      & " and we've made real progress."   \\ \midrule
50.44   &  53.06      & " Now it's exhilarating to me to think that as president,"   \\ \midrule
53.2  &  55.76      & " I could help to change all our people's lives for the better"   \\ \midrule
55.76    &  58.16      & " and bring hope back to the American dream."   
     \\ \bottomrule
\end{tabular}
    \caption{Example transcript and automatically detected text segments (and their corresponding start and end timestamps) extracted from Clinton's ca. $1992$ ``Hope, Arkansas'' ad (ad ID P-1291-61062).}
\label{tab:transcriptsegments}
\end{table}

Importantly, transcription with Whisper returns two outputs: the video transcripts and the segmentation of the transcript into short phrases that are separated by a natural pause in the speaker's voice or a change in the person speaking. Table~\ref{tab:transcriptsegments} presents an example of the transcription output.  Each text segment comes with the start and end points recorded in seconds, which we later combine with other information extracted from the video to determine the trimming start and end times.

\subsubsection*{Step 2: Video key frame extraction}

Our second task is to extract an ordered set of key frames (i.e., images of video frames) from each video that are \textit{comprehensive} in terms of reflecting each video's visual contents and progression of imagery. More specifically, we seek a time-ordered set of frames that contains all of the important visual content of the ad. 

Recent work has accomplished this by randomly sampling one frame for every 15 seconds of video, as well as one frame from the first and last two seconds of each ad  \cite{fowler2021politicalonlineoffline}.  However, our objective differs: rather than merely collecting a representative subset of frames, we aim to extract a frame sequence that comprehensively covers all visually relevant content in the ad.

We accomplish this through two complementary strategies. First, we determine when important images are shown according to the ad's narrator or speaker via to the transcription segments detected in \textbf{Step~1} above. Specifically, we observe that the visual contents of a particular ad are almost always closely time-synced to its transcript. For example, the vast majority of scene changes correspond to transitions between speech segments, which we detected in \textbf{Step~1} above.  Moreover, most visually important imagery is onscreen around the middle of a narrator's (or speaker's) speech segment. Therefore, for each video, we extract a frame corresponding to the central moment of each speech segment. In the example shown in Table~\ref{tab:transcriptsegments}, we extract the frame that is depicted at the timestamp that falls in the middle of the phrase ``I was born in \dots'' (i.e. the frame at $(4.4-0)/2 = 2.2$ sec), another frame from the middle of the phrase ``I remember that \dots'' (i.e. the frame at $6.97$ sec), and so on. 

However, it is possible that relying only on speech segments to detect important images may miss some important frames. This is because a small number of ads also avoid speech during key segments for dramatic effect or avoid spoken words entirely. For example, Clinton ran several ads that depict only text set to an upbeat jazz backing track. We account for these rare but important stylistic choices by extracting a `transcription-agnostic' set of video frames at regular $3$-second intervals. We then merge this set with the transcription-directed video frames described above.

\begin{figure}[t]
\centering
\includegraphics[width=0.43\textwidth]{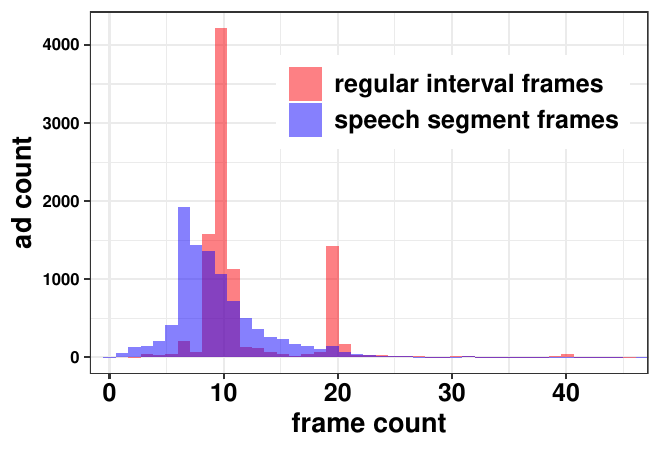} 
\caption[Framecounts]{Distributions of counts of frames extracted from ads in \textbf{Step 2}.}
   \label{fig:framehistogram}
\end{figure}

Fig.~\ref{fig:framehistogram} shows the distribution of counts of key frames extracted from the ads. The ads have a mean of $9.78$ text segment-centered frames plus $12.24$ regular-intervaled frames ($\sim$$22$ frames total). Note that in this figure, many ads have approximately $10$ or $20$ regular interval frames, as this corresponds to standard ad spot lengths of $30$ or $60$ seconds.

\subsubsection*{Step 3: Obtaining video key frame descriptions}

We generate a brief summary of each video key frame by submitting it to a multimodal LLM: \textsc{GPT-4-vision}, accompanied by a custom prompt that provides context using the ad metadata. Specifically, we submit the following prompt for each key frame:

\begin{quote}
\textit{Describe what is depicted in this video frame in no more than $15$ words. Do not state that the frame depicts a vintage advertisement, and do not comment on the image quality. If the image includes text, then state that it includes text and also include a summary of the text that is shown. For context, this video frame is a still taken from an advertisement for the \{ELECTION YEAR\} presidential campaign of \{PARTY\} \{CANDIDATE\}. The transcript of the entire ad is: \{TRANSCRIPT\}
}
\end{quote}

In some cases, attack ads avoid advocating for a specific candidate, and instead focus entirely on criticizing an opponent. For these cases, we replace the phrase \textit{``\dots an advertisement for the \{ELECTION YEAR\} presidential campaign of \{PARTY\} \{CANDIDATE\}\dots''} with the phrase \textit{``an advertisement for the \dots\{ELECTION YEAR\} presidential election. This ad is anti-\{CANDIDATE\} and pro-\{PARTY\}\dots''}

\subsubsection*{Step 4: Video summarization}

We now describe how we obtain LLM-generated summaries of the ads. Specifically, we generate each ad summary by querying GPT using a customized summarization prompt that provides (1) the ad transcript generated in \textbf{Step~1}; (2) the time-ordered set of approximately 20 descriptions of video imagery generated in \textbf{Step~3} above including the merged speech-segment-based key frames and regular interval frames (merged by ascending timestamps); and (3) a contextual description of the ad that contains relevant ad metadata. We consider the following prompt, which we submit to the GPT-4 API:

\begin{quote}
\textit{Provide a $50$ word summary of a political television ad for the academic community. Your summary should not exceed $50$ words. For context, this ad was for the \{ELECTION YEAR\} presidential campaign of \{PARTY\} candidate \{CANDIDATE\}. The transcript of the entire ad is:\\
     \{TRANSCRIPT\}\\
     The ad video depicts a set of scenes that can be described as follows:\\
     \{First ad frame description\}\\
     \{Second ad frame description\}\\
     \dots
}
\end{quote}
 
We note that our prompt repeats its emphasis on the 50-word length limit. This is by design: during testing, we noted that including only a single reference to the $50$-word limit in the prompt resulted in summaries that exceeded $50$ words over $10\%$ of the time, whereas the doubly-repeated emphasis results in $50$-word summaries in nearly all cases. We also note that when ads criticize a candidate rather than advocate for one, we apply the same prompt modification (context phrase modification) as we did for its frame description prompts above.

\paragraph{Example summary output} Our method returns the following ad summary for the Hope, Arkansas example ad: \textit{"The 1992 political ad features then-candidate Bill Clinton narrating a touching dialogue of his humble beginnings in Hope, Arkansas, his work in public service, and his dreams of making a difference as President. Visually, the ad is a blend of his past and present, reflecting a life dedicated to civic engagement."}

\subsection*{Scaling up to large video datasets}
\label{ssec:scalingup}

We design our codebase to accommodate even very large-scale applications (including those that are orders of magnitude larger than the project described here) by using the MPI \cite{walker1996mpi} parallelization standard via \textsc{Open MPI} \cite{gabriel2004open} and the \textsc{MPI4py} \cite{dalcin2021mpi4py} Python module. Specifically, we note that the workflow described above exhibits what is known as `embarrassingly parallelizability' at the level of videos (i.e. each video can be computed fully independently in parallel). Key frame description generation (\textbf{Step~3}) can also be parallelized at the level of individual ad frames. We conduct all processing steps described above by computing on approximately $50$ workloads simultaneously, which is sufficient for our purposes. However, we note that our codebase and techniques may easily be applied to much larger datasets (e.g., millions of videos) \textit{without} requiring additional compute time simply by scaling up the processor count.

\section*{Data Records}

All presidential ads and associated transcripts, summaries, key frames, and metadata are available for \emph{bulk download} on Harvard's Dataverse at: \url{https://dataverse.harvard.edu/dataverse/ctap} \ \citep{ctap2025}.  We have organized the bulk data into subdirectories with one directory for each election year so that scholars interested in a particular subset of elections may bulk-download just that subset of interest. The filenames for each transcript, summary, and ad key frame begin with the filename of the video to which they correspond. That video filename also matches a corresponding row in the dataset metadata file that includes the ad's metadata. In the \emph{Usage Notes} section below, we also describe our online interface designed to allow researchers to efficiently explore the data. 

\section*{Technical Validation I: Data Accuracy, Consistency, and Balance versus Human Coders}
\label{sec:humvalidation}

Our goal in this validation section is to show that (1) our workflow produces high-quality preprocessed videos,  transcripts, and summaries \textit{overall}, and (2) the quality is \textit{consistent across features of interest}. We accomplish this by showing that across these three tasks, our methodology performs at least as well as human coders using standard measures of quality and this performance is independent of features such as partisanship and year.

\subsection*{Stratified Random Sampling}

Our dataset is highly diverse, spanning not only competing political parties, but also seven decades of changes to speech styles and issue areas under discussion. The dataset also spans the revolution in audiovisual technology from black-and-white video cameras with mono vinyl-cut audio to high-definition color video with digitally processed audio. To ensure the integrity of downstream research, it is advantageous to explicitly test whether any of these sources of variance (including aspects that are imperceptible to human senses) cause our AI-based workflow to exhibit errors that affect one party or a certain time range of videos more than others. 

To accomplish this, we employ a stratified sampling validation strategy. Specifically, instead of validating uniformly randomly sampled ads, we instead validate by sampling batches, where each batch contains one sampled ad for each party-year combination. For tasks that required additional statistical power, we asked our human research assistants to label additional batches of stratified samples. As a result of this sampling strategy, the high performance on the validation tests below indicates high quality that is \textit{balanced} across the primary sources of variation in the dataset. We also test this explicitly by showing that quality has no statistically significant relationship with features of interest.

\subsection*{Preprocessing Validation}
We empirically validate our preprocessing procedure that trims pre-roll and post-roll clips out of each ad's video file.
Three undergraduate research assistants were asked to manually identify the beginning and end of political ads for $30$ stratified sample batches of ads (i.e., $900$ video files---see the ``Stratified Random Sampling'' section above). We then compare these human-coded ad beginning and ending timestamps with those inferred by our automated procedure.

\paragraph{Quality Evaluation Metrics.}
Our pipeline already ensures that no dialogue or onscreen people are `overtrimmed' from the video, and that any pre-dialogue or post-dialogue segments of the ad will also be sampled and included in our AI summarization prompt via our regular-spaced $3$-second interval key frame samples. Therefore, our primary goal in this section is to show that our automated procedure removes `pre-roll' and `post-roll' without overtrimming more than $3$ seconds of pre-dialogue or post-dialogue imagery from the ads. 

\paragraph{Preprocessing Evaluation Results.}
Compared to our human coders, our automated preprocessing procedure `overtrims' the start (`pre-roll') of the videos by more than $3$ seconds in approximately $2.44$\% of the stratified-sampled ads.  Similarly, compared to our human coders, our automated preprocessing procedure `overtrims' the end (`post-roll') of the videos by more than $3$ seconds in approximately $3.33$\% of the stratified-sampled ads. This ensures that our automated preprocessing step obtains high accuracy for this task. 

We also measure cases where our automated procedure `undertrims' the beginning or end of the ads (i.e. leaves additional video frames) compared to humans, though such `errors' are generally benign. Specifically, our automated procedure `undertrims' the videos `pre-roll' by more than $3$ seconds in approximately $3.44$\% of the sampled ads, and it `undertrims' the `post-roll' in approximately $12.67$\% of the sampled ads. In cases where there is 'undertrimming' of the end, the mean undertrim leaves $2.38$ seconds of additional video compared to human coders.

We also explicitly test whether errors affect one party or a certain time range of videos more than others. Specifically, we regress the difference between human-coded times and automated times for each ad on party, election year, and a party-election-year interaction term by estimating models of the form:
\begin{align}
    \text{Error}_{type} = \beta_0 + \beta_1 \text{Republican} + \beta_2 \text{ElectionYear} + \beta_3 \text{Republican}\times \text{ElectionYear}
    \label{beta-preproc}
\end{align}
We refit this model for different error types including ($1$) the videos' start time errors, ($2$) the videos' end time errors, and ($3$) the videos' sum of absolute errors, i.e.,$\text{Error}_j = |\text{Error}_{j,start}| + |\text{Error}_{j,end}|$. In all cases, non-intercept coefficients $\beta_1, \beta_2, \beta_3$ are statistically insignificant ($p$-value $>0.11$). This indicates that errors compared to human coders have no statistically significant relationship to these features of interest. We report details in the Appendix.

We also test the intercoder reliability of our three undergraduate research assistants on this task. Specifically, we uniformly subsampled a subset of $\sim$$200$ of these sampled ads and asked all three research assistants to identify start and end timestamps for these ads. We then compute Intraclass Correlation Coefficient (ICC) for the raters using the appropriate two-way random effects, absolute agreement, single rater/measurement model following the standard approach \cite{mcgraw1996forming}. Our three raters have an ICC of $0.97$ for their start times and $0.98$ for their end times, indicating excellent inter-coder agreement. More concretely, across the stratified samples, the mean absolute difference between any two human coders' is $0.62$ seconds for `pre-roll' timestamps and $0.67$ seconds for `post-roll' timestamps. This ensures that our human-generated timestamps provide a high-quality standard of comparison.

\subsection*{Automated Transcription Validation}
Our ad transcripts from \textbf{Step 1} are generated by OpenAI's \textsc{whisper-large-v3} model, which has already been extensively validated in other contexts \cite{radford2023robust}. Therefore, our goal is just to ensure that its performance on our ads is consistent with its known performance on benchmark video datasets. To accomplish this, we ask our human coders to manually transcribe a smaller stratified sample of $4$ stratified batches of sampled ads (i.e., $120$ ads---see ``Stratified Random Sampling'' Section above).

\paragraph{\textbf{Quality Evaluation Metrics.}}
The standard evaluation metric used to compare the automated transcripts to manually generated transcripts is Word Error Rate (WER). WER represents the count of errors in the AI transcript divided by the word count $n$ of the reference (correct) transcript \cite{hamed2023benchmarking}. Here, an error is defined as either a substituted word (denoted by $s$), a deleted/missing word (denoted by $d$), or a superfluously inserted word (denoted by $i$). Thus, we compute WER as:
\begin{ceqn}
\begin{align}
    \text{WER} \coloneq \frac{s+d+i}{n}\end{align}
\end{ceqn}

A WER of $0$ indicates that the candidate text (i.e., the automated transcript) and the reference text (i.e., human transcript) are identical, while a WER of $0.10$, for example, indicates that there are $10$ errors for every $100$ words in the reference transcript. For reference, the original Whisper paper \cite{radford2023robust} showed state-of-the-art WER's across $14$ standard benchmark datasets ranging from a WER of $0.027$ on `maximally clean' high-definition speech data recorded in e.g. sound isolating voiceover booths (\textit{LibriSpeech Clean} data) to a WER of $0.364$ on speech audio recorded in meeting rooms (\textit{AMI SDM1} data). In general, state-of-the-art transcription obtains WER values of $0.16$--$0.26$ on standard `in-situ' speech benchmark datasets \cite{radford2023robust}. For comparison, a single human transcriber has a WER of approximately $0.14$ on these same benchmark datasets. 

\paragraph{\textbf{Transcription Validation Results.}} We compute two WER metrics for our validation. First, we compute the WER for each transcript and then take the mean over the transcripts. The transcripts have a mean WER of $0.0353$. Second, we also consider the overall WER, which we compute by concatenating the $120$ transcripts and calculating a single overall WER. This yields an overall WER of $0.0329$. In both cases, we conclude that our transcripts exhibit roughly $3$ errors per $100$ words. This means that our transcripts exhibit higher accuracy than standard benchmark transcriptions of virtually all standard transcription datasets except for `maximally clean' high-definition speech data recorded in voiceover booths (e.g. transcriptions of \textit{LibriSpeech Clean} data obtain WER that differs from ours by a margin of less than a single error in every $100$ words \cite[see][]{radford2023robust}).

We also find that these rare transcription errors are qualitatively negligible in terms of their effect on the meaning of the text. Table~\ref{tab:transcripterrors} shows $5$ transcripts randomly selected from the subset of validation transcripts that exhibited any errors. Note that errors are generally limited to minor spelling differences or pronouns, and do not change the substantive meaning of the text.

\begin{table}[t]
\centering
\tabcolsep=3.4pt
\small
\begin{tabular}{@{}|p{1.01\linewidth}|@{}}
\toprule\toprule
\textit{ID: P-1232-57589 (WER: 0.0494)}: "For Hispanics like me, \textit{\st{this present}}[\textbf{discretion in}]  election\textbf{s} may be the most important in our history because Hispanics everywhere have a real chance to be heard, to elect a president who cares about our needs, our proud traditions, our strong family values. But don't take my word for it. Ask my father-in-law. As president, I will have a lot of reason\textbf{\st{s}} to help Hispanics everywhere because I'll not only be answering to my grandchildren, I will be answering to history."\\
\midrule
\textit{ID: P-1905-108595 (WER: 0.0133)}:
"The people of Florida face an important choice do \textit{\st{you}}[\textbf{we}] elect a president who will raise your taxes, or fight for working families? A president who will spread your income, or let you keep what is yours? A president who will spend us deeper into debt, or control spending and eliminate waste? Tough times take experienced bipartisan leadership. Join me and vote for senator John McCain. I am John McCain and I approve this message."\\
\midrule
\textit{ID: P-334-15180 (WER: 0.109):} \textit{\st{whose}}\textbf{[who is]} bill did Congress pass to keep the lid on oil prices? Udall Bayh S\textit{\st{h}}[\textbf{c}]hriver Wallace Sanford      \textit{\st{Shapp}}[\textbf{Schaap}]  Carter Harris? Nope, Senator Scoop Jackson. Who's the only candidate to win the Sierra Club's highest environmental award? Wallace Sanford  \textit{\st{Shapp}}[\textbf{Schaap}] Harris Carter S\textit{\st{h}}\textbf{c}hriver Bayh Udall? Nope, Senator Scoop Jackson. Which candidate insists \textit{\st{that}}  we stop exporting our jobs to Russia? Sanford  \textit{\st{Shapp}}[\textbf{Schaap}] Harris Wallace Udall Carter S\textit{\st{h}}[\textbf{c}]hriver   Bayh? Nope, Senator Scoop Jackson. Let's have a president who does something. Elect Senator Scoop Jackson while there is still time to save the 70's.\\
\midrule
\textit{ID: P-1631-83508 (WER: 0.0118):} 
"As Governor, George W. Bush gave big oil a tax break, while opposing health care for 220,000 kids. Texas now ranks 50th in family health care. He's left the minimum wage at \$3.35 an hour, let polluters police themselves. Today, Texas ranks last in air quality. Now Bush promises the same one trillion \textit{\st{dollars}} from Social Security to two different groups. He squanders the surplus on a tax cut for those making over \$300,000. Is he ready to lead America?" \\
\midrule
\textit{ID: P-1840-104205 (WER: 0.0108):} 
"There's nothing wrong with making a decision and then changing your mind. But if you never commit to what you believe in, who will ever commit to you? John Kerry has changed his mind on all these important issues. Now, there's nothing wrong with a little indecision as long as your job doesn't involve any responsibility. John Kerry has changed his mind time and time again. If you thought you could trust him, you might want to change your mind too. Clubfo[\textbf{u}]rGrowth.net is responsible for the content of this advertising."\\ 
\bottomrule \end{tabular}\caption{\textbf{Transcription error examples.} for five  transcripts randomly selected from validation transcripts with nonzero errors. Words in the reference transcript that the AI transcript missed are denoted by strikethrough text. Words (or single letters) that the AI transcript erroneously added are denoted by bold text.}
\label{tab:transcripterrors}
\end{table}
 
Finally, we test whether transcription errors affect one party or time range of ads more than others. Specifically, we model transcription accuracy as a function of party, election year, and a party-election-year interaction term. Here, we are careful to adopt modeling choices that reflect the word-error distributions of the transcripts. In particular, it is undesirable to model word errors directly, as errors within the same transcript are not independent. It is also undesirable to adopt WER as the response, as WER has a lower bound ($0$) but no upper bound in the presence of insertions, and the data-generating process for WER is unclear with respect to insertions.

Therefore, we consider a transcript-level response variable that is bounded and linear in the count of per-transcript errors. Specifically, we define a response variable, $R_j$, that captures transcription accuracy as the fraction $0 \le R_j \le 1$ of words in a human transcript $j$ that are correctly recovered by the AI transcript: $$R_j \coloneq \frac{n_{i} - s_j - d_j}{ n_{i}}.$$ We then use Beta regression to model this fractional response. Because the logit link used in the Beta regression is undefined when the response takes endpoint values, we adopt the standard approach of \cite{smithson2006better, smithson2006bettersupp} and squeeze response values by a negligible amount: $R_j \in [0,  1] \rightarrow R'_j \in (0,  1)$, by defining $R'_j \coloneq (R_j (N -1) + 0.5) / N$, where $N$ is the sample size (i.e. $120$). Now, we can proceed with the standard Beta regression where the transcription accuracy (response variable) $R'_j$ follows a beta distribution with mean $\mu_j$ and precision $\phi$, where the mean is linked to covariates via the standard logit link function $g(x)$:
\begin{align}
    \label{beta-transcript}
    R'_j \sim \text{Beta}(\mu_j \phi, (1 - \mu_j) \phi), \quad\quad g(\mu_j) = \beta_0 + \beta_1 \text{Republican}_j + \beta_2 \text{ElectionYear}_j + \beta_3 \text{Republican}_j\times \text{ElectionYear}_j
\end{align}
Once again, we find that all non-intercept coefficients, $\beta_1, \beta_2, \beta_3$, are statistically insignificant (all $p>0.5$), indicating that the high transcription accuracy we observe is consistent regardless of party and year. We defer details to the Appendix. 

\subsection*{Video Summarization Validation}
The goal of our AI-generated ad summaries is to facilitate academic research on political campaign ads. As such, our workflow should provide researchers with a coherent, neutral, accurate summary that efficiently conveys the main themes and relevant or interesting aspects in each ad. Therefore, we seek summaries that exhibit high quality across \textit{multiple} key dimensions. We note that this goal differs in some important respects from the objective functions used to train standard LLMs, which emphasize only fluency (via token prediction accuracy).

\paragraph{\textbf{Quality Evaluation Metrics.}} To measure these dimensions, we ask human subject matter experts to evaluate the summaries based on the $4$ quality evaluation criteria of \cite{kryscinski2019neural} and \cite{fabbri2021summeval}: \textit{coherence, consistency, fluency,} and \textit{relevance}. Table~\ref{t:defs} gives the definitions of each of these criteria. 

\paragraph{\textbf{Comparison Benchmark: Human-Generated Summaries.}} We compare a validation set of our $50$-word AI-generated summaries from \textbf{Step 4} to $50$-word summaries composed by undergraduate research assistants (we note that in recent work on political ads, most ad metadata is produced by undergraduate research assistants). Specifically, for each ad in $4$ stratified sample batches (i.e. $120$ sampled ads), we randomly selected one of our undergraduate RA's and asked them to compose a summary via the following prompt: \\

\begin{quote}
\textit{We are releasing a large collection of political ads to the academic community.  Summaries of these ads will also be released.  Please help this project by writing a $50$-word summary of the ad you have been assigned.  The summary should be written in a way that is useful for academic research. Also, since we know the names of the candidates, please do not write their names in your summary. Rather, just refer to them as `the candidate' or `the opposing candidate.'}
\end{quote}

\paragraph{\textbf{Expert Human Annotators and Online Evaluation Setup.}} To ensure fairness in the evaluation, we created an online evaluation tool using a JavaScript embedded within a Qualtrics survey. This tool randomly displayed an ad and two summaries, labeled ``Summary A'' and ``Summary B." The human summary was randomly assigned to either ``Summary A'' or ``Summary B,'' and the LLM-generated summary assigned the remaining label. We then asked political science graduate students to answer four questions about each summary, corresponding to the four summary evaluation metrics identified in the literature: coherence, consistency, fluency, and relevance \cite[][396]{fabbri2021summeval}. For example, to evaluate a summary's consistency, which is defined as the ``factual alignment between the summary and the summarized source,'' we asked to what extent someone agreed or disagreed with the following statement: ``The summary is factually consistent with the advertisement.'' Annotators could select from $5$ responses ranging from ``strongly disagree'' to ``strongly agree'', which we coded on a 5-point integer Likert scale as $-2$ and $2$, respectively. We ran the evaluation until each ad in the stratified sample had been rated by three graduate students.

\paragraph{\textbf{Summary Validation Results.}} 
We employ a range of statistical methods to analyze the comparative quality of human- and AI-generated ad summaries. Table~\ref{t:summary} reports mean responses of three graduate research assistants across the four quality dimensions. We include standard paired $t$-tests as well as Wilcoxon signed-rank tests; the latter tests are more robust given that our response variables are  Likert-scale ratings. According to both types of tests, the mean score of our AI summaries taken across all quality dimensions outperforms the human summaries by a statistically significant margin (Table~\ref{t:summary}, bottom row). Our AI summaries also exhibit significantly better fluency---that is, they are more likely to convey the main themes of each ad. Other dimensions exhibit either a slight advantage for our AI summaries, or no statistically significant difference between AI and human summaries. 

While paired $t$-tests and Wilcoxon signed-rank tests provide valuable baseline comparisons, they do not account for within-rater dependencies or control for key features such as year and party. To address these aspects, we also estimate a Mixed-Effects Ordinal Logit (MEOL) model, a form of Cumulative Link Mixed Model that appropriately models the ordinal (Likert) dependent variable while incorporating random effects to account for dependencies in the data. Our MEOL model includes:

\begin{itemize}
  \setlength{\itemsep}{2pt}  \setlength{\parskip}{0pt}  \setlength{\topsep}{0pt}   \item \emph{Treatment effect of AI vs. human authorship} ($\beta_1$) measures the causal effect of AI-generated summaries on ratings;
\item \emph{Covariates for year, party, and their interaction} ($\beta_2$, $\beta_3$, $\beta_4$) control for systematic differences in ads;
  \item \emph{Random intercepts for raters} ($u_{\text{Rater}}$) capture latent differences in individuals' scoring tendencies (within-rater correlation);
  \item \emph{Random intercepts for ads} ($u_{\text{Ad}}$) account for different ads' latent variation in summary difficulty.
\end{itemize}

Specifically, the MEOL model estimates the probability that, for one of the four quality dimensions $d$, a given summary rating $Y$ falls within a particular Likert category. The MEOL model takes the form:
\begin{equation}
\text{logit} \left( P(Y_d \leq k) \right) = \tau_k - \left( \beta_1 \text{AI} + \beta_2 \text{Republican} + \beta_3 \text{ElectionYear} + \beta_4 \text{ElectionYear} \times \text{Republican} \right) + u_{\text{Rater}} + u_{\text{Ad}}
\end{equation}

\bgroup
\begin{table}[t]
\caption{\textbf{Dimensions of summary quality rated by human coders.} }\label{t:defs}
\centering
\small
\begin{tabular}{|ll|}
  \toprule Dimension & Definition \\ 
   \midrule
   \midrule
Coherence & Coherent and easy to understand.       \\ 
   Consistency & Factually consistent with the advertisement.      \\ 
   Fluency & Contains the main themes of the advertisement.         \\ 
   Relevance & Useful for research on presidential advertisements.\\ 
\hline
\bottomrule
\end{tabular}
\end{table}
\egroup
 
\noindent Here, $Y$ represents a Likert rating of a summary in one of the quality dimensions $d$, and $P(Y \leq k)$ denotes the cumulative probability of receiving a rating in category $k$ or lower. Note that the negative sign before the predictor effects means that larger values of $X\beta$ decrease the cumulative probability $P(Y_d \leq k)$ of a lower rating, shifting the probability mass toward higher ordinal categories. This matches our intuition that larger $X\beta$ corresponds to a better rating.
The thresholds $\tau_k$ then define the latent decision boundaries between ordinal categories. The terms in parentheses capture the main predictors of interest that estimate the treatment effect of AI summaries on ratings and control for covariates such as party and year. Finally,  $u_{\text{Rater}}$ and $u_{\text{Ad}}$ terms are random intercepts that account for individual-level variation in rater tendencies and latent characteristics of the ad summaries, respectively. By including both random effects ($u$-terms) and covariates, the MEOL model allows us to properly account for the pairwise structure of the experiment, where each rater evaluates both the human- and AI-generated summary for the same ad, while also controlling for individual rater biases and latent ad characteristics that could influence ratings.

\bgroup
\addtolength{\tabcolsep}{-0.4em}
\begin{table}[t]
\caption{\textbf{Video summarization validation results.}  AI-generated summaries perform at least as well as human summaries. $t$-statistics are from 2-sided $t$-tests. $V$-statistics are from paired samples Wilcoxon tests ({Note:}  $^{*}$p$<$0.1; $^{**}$p$<$0.05; $^{***}$p$<$0.01). The sample size of $N=120$ paired ratings per quality dimension corresponds to $120$ stratified-sampled movies, each with one AI-generated summary and one human-written summary. Each summary was rated by three independent raters, and their ratings were averaged to produce a single score per summary per quality dimension.}\label{t:summary}
\centering
\small
\begin{tabular}{|lccc|lll|ll|}
  \toprule Dimension  & AI mean & Human mean & diff. & $t$-test 95\%  C.I. & $t$-statistic  & $p$-value & $V$-statistic  & $p$-value \\ 
   \midrule
   \midrule
&&&&&\\[-1.8ex] 
 Coherence &  1.52 & 1.37 &0.15 & (-0.018, 0.313)    & 1.753 & 0.08$1^*$                & 1493.5   &   0.104\\ 
   Consistency & 1.54 & 1.61 & -0.07 & (-0.231, 0.087) & 0.895 & 0.371                 & 1856.0   &   0.491\\ 
   Fluency & 1.64 & 1.49 &0.15 & (\phantom{-}0.005, 0.289) & 2.041  & 0.04$2^{**}$    & 1261.5   &   0.02$8^{**}$ \\ 
   Relevance & 1.51 & 1.43 &  0.08 & (-0.075, 0.242) & 1.037 & 0.301                   & 1609.5   &   0.146\\ 
  &&&&&\\[-1.8ex] 
   \hline
    &&&&&\\[-1.8ex] 
Aggregate  ($N$$=$$480$ pairs)  & 1.55 & 1.48  &   0.08 & (-0.001, 0.153) &  1.956  & 0.05$2^{*}$  & 2258.5   &  0.00$9^{***}$\\
  \bottomrule
\end{tabular}
\vspace{-1em}
\end{table}
\egroup

\bgroup
\addtolength{\tabcolsep}{-0.7em}
\begin{table}[!htbp] \centering 
\vspace{0.3em}
  \caption{Human vs. AI Summary Results: Mixed-Effects Ordinal Logit (MEOL), OLS, \& OLS with ad fixed effects (OLS-FE). SE's in parentheses; bootstrapped SE's in brackets; OLS and OLS-FE SE's are robust \& ad (OLS) or rater (OLS-FE) clustered.}
  \label{tab:meol} 
\begin{tabular}{@{\extracolsep{5pt}}l l l l l l l l l l l l l}
\\[-1.8ex]\hline 
\hline \\[-1.8ex] 
 & \multicolumn{3}{c}{Coherence} & \multicolumn{3}{c}{Consistency} & \multicolumn{3}{c}{Fluency} & \multicolumn{3}{c}{Relevance} \\ 
\cmidrule{2-4} \cmidrule{5-7} \cmidrule{8-10} \cmidrule{11-13}
\\[-1.8ex] & \multicolumn{1}{|c}{\emph{MEOL}} & \multicolumn{1}{c}{\emph{OLS}} & \multicolumn{1}{c|}{\emph{OLS-FE \  \ }} & \multicolumn{1}{c}{\emph{MEOL}} & \multicolumn{1}{c}{\emph{OLS}} & \multicolumn{1}{c|}{\emph{OLS-FE \  \ }} & \multicolumn{1}{c}{\emph{MEOL}} & \multicolumn{1}{c}{\emph{OLS}} & \multicolumn{1}{c|}{\emph{OLS-FE \  \ }} & \multicolumn{1}{c}{\emph{MEOL}} & \multicolumn{1}{c}{\emph{OLS}} & \multicolumn{1}{c|}{\emph{OLS-FE \  \ }}\\ 
\hline \\[-1.8ex] 
AI              & \phantom{(}0.335\textsuperscript{**}    & \phantom{(}0.147\textsuperscript{*}  & 0.147\textsuperscript{***} & -0.171  & -0.072     & -0.072 & \phantom{(}0.620\textsuperscript{**}       & \phantom{(}0.147\textsuperscript{**}    & 0.147\textsuperscript{***} & \phantom{(}0.332\textsuperscript{**}     & \phantom{(}0.083 & 0.083 \\ 
& (0.164)       & (0.078)    & (0.055)  & (0.186) & (0.072)    & (0.056)  & (0.177)          & (0.069)       & (0.052)  & (0.165)        & (0.073) & (0.055) \\ 
                 & \textcolor{gray}{[0.179]} &                  &                  & \textcolor{gray}{[0.194]} &            &            & \textcolor{gray}{[0.193]}          &               &               & \textcolor{gray}{[0.179]}        & \\ 
                   
  & & & & & & & & & & & & \\ 
Republican      &  \phantom{(}0.300       & -2.655 &       & \phantom{(}0.341    & \phantom{(}4.280  &   & \phantom{(}0.317       & -5.917   &    & \phantom{(}0.236      & \phantom{(}8.417 &  \\ 
& (0.278)      & (11.035)   &   & (0.283)  & (10.921)  &   & (0.265)      & (8.901)  &   & (.265)   & (10.407) &  \\ 
                 & \textcolor{gray}{[0.226]} &                 &             & \textcolor{gray}{[0.233]} &            &           & \textcolor{gray}{[0.216]}          &           &           & \textcolor{gray}{[0.218]}        & \\ 

  & & & & & & & & & & & & \\ 
ElectionYr       & \phantom{(}0.013      & \phantom{(}0.002   &   & \phantom{(}0.020\textsuperscript{*}      & \phantom{(}0.007\textsuperscript{*}  &  & \phantom{(}0.008      & \phantom{(}0.0004  &   & \phantom{(}0.019\textsuperscript{*}      & \phantom{(}0.006 & \\ 
& (0.011)    & (0.004)  &   & (0.011)    & (0.004)        &   & (0.010)     & (0.003)       &   & (0.010)       & (0.004) &  \\ 
                    & \textcolor{gray}{[0.009]} &             &   & \textcolor{gray}{[0.010]} &                   &   & \textcolor{gray}{[0.009]}          &          &          & \textcolor{gray}{[0.009]}        & \\  
  & & & & & & & & & & & & \\ 
Rep.*El.Yr. & -0.008       &\phantom{(} 0.001   &   & -0.002     & -0.002     &   & \phantom{(}0.006       & \phantom{(}0.003   &   & -0.021     & -0.004 & \\ 
& (0.016)      & (0.006) &   & (0.016)    & (0.005)    &   & (0.015)     & (0.004)    &   & (0.015)    & (0.005) & \\ 
                  & \textcolor{gray}{[0.014]} &              &   & \textcolor{gray}{[0.014]} &               &   & \textcolor{gray}{[0.013]}          &          &          & \textcolor{gray}{[0.013]}        & \\ 
  & & & & & & & & \\ 
\hline \\[-1.8ex] 
Observations & \multicolumn{1}{c}{720} & \multicolumn{1}{c}{720} & \multicolumn{1}{c}{720} & \multicolumn{1}{c}{720} & \multicolumn{1}{c}{720} & \multicolumn{1}{c}{720} & \multicolumn{1}{c}{720} & \multicolumn{1}{c}{720} & \multicolumn{1}{c}{720} & \multicolumn{1}{c}{720} & \multicolumn{1}{c}{720} & \multicolumn{1}{c}{720} \\ 

\hline 
\hline \\[-1.8ex] 
\textit{Note:}  & \multicolumn{12}{r}{$\textsuperscript{*}$p$<$0.1; $\textsuperscript{**}$p$<$0.05; $\textsuperscript{***}$p$<$0.01} \\ 
\end{tabular} 
\end{table} 
\egroup

Table~\ref{tab:meol} reports the results of the $MEOL$ regression computed for each of the four quality dimensions, Coherence, Consistency, Fluency, and Relevance. To assess statistical significance, we report both standard errors (in parentheses) and ad-clustered bootstrapped standard errors (in square brackets). While MEOL does not assume normality of residuals in the way that linear regression does, it nonetheless relies on the assumption that random effects are normally distributed. Therefore, bootstrapped standard errors allow us to test for significant effects even under potential violations of these assumptions. 

Finally, we also estimate two variants of OLS linear regression for each quality dimension $d$ and report it alongside the corresponding MEOL estimates. First, we estimate an OLS model with party, election year, and interaction controls as above. We report standard cluster-robust errors clustered by ad to account for potential within-ad rating correlations:
\begin{equation}
\text{\emph{OLS:\ \ \ \ }}Y_d = \alpha + \beta_1 \text{AI} + \beta_2 \text{Republican} + \beta_3 \text{ElectionYear} + \beta_4 (\text{ElectionYear} \times \text{Republican}) + \epsilon,
\end{equation}

Second, we estimate an OLS model with ad fixed effects (\emph{OLS-FE}), which fully absorb party and year controls, and we report the usual Huber-White standard errors. This Neymanian approach allows us to explicitly restrict measurement of the AI treatment effect to pairwise AI-vs-human summary comparisons:
\begin{equation}
\text{\emph{OLS-FE:\ \ \ \ }}Y_d = \beta_1 \text{AI} + \gamma_{ad} + \epsilon,
\end{equation}
While OLS models do not account for the ordinal nature of the Likert ratings or the hierarchical structure of our data, they nonetheless provide useful and familiar baselines.

\paragraph{\textbf{MEOL and OLS Results of Human vs AI Summaries.}} 
We find that AI summaries exhibit significantly better coherence, fluency, and relevance compared to their human-authored counterparts. Specifically, MEOL coefficients represent the log-odds of moving into a higher category of the ordinal response (i.e. rating). Therefore, we exponentiate these coefficients to obtain interpretable results. We find that our AI summaries have:
\begin{itemize}
  \setlength{\itemsep}{2pt}  \setlength{\parskip}{0pt}  \setlength{\topsep}{0pt}   \item 39.8\% greater odds of obtaining a higher coherence score;
    \item 85.9\% greater odds of obtaining a higher fluency score;
    \item 39.4\% greater odds of obtaining a higher relevance score.
\end{itemize}

The remaining dimension, consistency, is the only dimension where the AI-generated summary was not rated significantly higher than the human-generated summary, suggesting that both are equally adept at capturing the factual information in the ad. We also find that the remaining covariates are statistically insignificant, with the exception of election year, which is moderately correlated with two of the four quality dimensions at the $p<0.1$ level. We note that the magnitude of these potential time effects remains small. 

\section*{Technical Validation II: Tracing the Historical Emergence of Key Issues}

We briefly validate and illustrate our dataset by considering an application that measures the historical evolution of current key electoral issue areas. A central focus of political communication research is understanding how such issues emerge and shift over time \cite{petrocik2003issue, hayes2005candidate, vavreck2009message, fowler2021politicalonlineoffline}. This task not only demonstrates the dataset’s substantive relevance, but also serves as a validation check: do the trends it reveals correspond to historically recognized developments in campaign issue agendas? We find that our primary topics are generally consistent---though not perfectly aligned---with a well-known coding of subjects and themes that analyzes the transcripts from a smaller set of $895$ campaign advertisement from $1952$-$2000$ \citep{vavreck2009message}. Such validation is especially valuable given the difficulty of analyzing electoral issue dynamics prior to the 1996 election due to limited data availability and high labor costs involved in coding more than a few hundred ads \citep{vavreck2009message}.

\begin{table}[t]
\centering
\small
\begin{tabular}{|lll|}
  \toprule Topic & Election & Keywords \\ 
   \midrule
   \midrule
Economy &  $2024$, $2020$, $2016$ & economy, employment, unemployment, inflation, growth, price, recession, markets, trade, wage \\
Immigration & $2024$  &  immigration, immigrant, border, naturalization, alien, asylum \\
  Abortion &  $2024$ \& advertiser  &  abortion, pro-life, pro-choice, roe, roe v wade, termination, maternal, \\
  Terrorism & $2020$  &  terrorism, isis, isil, homeland, radical, security, osama, laden, attack, 9-11, al-qaeda \\
  Foreign Policy & $2020$  &  nato, china, russia, benghazi, alliance, foreign, war, peace \\
  Supreme Court & $2016$  &  supreme, court, appointment, justices, ginsburg, barrett \\
  Healthcare &  $2016$ &  healthcare, obamacare, medicare, prescription, insurance \\
    Taxation       & advertiser & taxation, taxes, bracket, income \\
  Crime        & advertiser & crime, police, prison, sentence, criminal, gang, murder, rape, rapist, murderer, superpredator\\
   \bottomrule
\end{tabular}
\caption{\textbf{Keyword topics and their associated keywords.} Keywords for topics $1$ to $7$ represent the top $3$ issue areas by voter interest from the most recent $3$ elections according to Pew Research (see the \textit{Election} column). Keywords $8$ and $9$ on Taxation and Crime reflect two remaining issue areas that are top-$3$ issues (along with Abortion) per current advertiser attention.}\label{t:keyatm}
\end{table}

To accomplish this, we estimate a keyword-assisted dynamic topic model (dynamic keyATM) on the corpus of ad transcripts \cite{eshima2024keyword}. This model conveys two advantages for this application. First, the model allows topic proportions to evolve across election years.  Second, it permits us to seed keyword topics corresponding to the top issue areas from the three most recent presidential elections ($2024, 2020, 2016$). 

We select keywords for our dynamic keyATM model by choosing issue areas that reflect the three issues most important to voters in these three elections according to Pew Research \cite{pew2024, pew2020, pew2016}. Due to some overlapping issues across these elections, this yields a total of $7$ keyword-seeded topics, which we show in Table~\ref{t:keyatm}. However, according to recent research, campaign messaging often focuses on a set of topics that diverge from voters' priorities. For example, taxation and crime---two issues that do not appear in voters' top three priorities---are among the top three issue areas in the $2024$ election in terms of advertiser attention \cite{AdImpactReview, AdImpactPlay}. We therefore include two additional keyword-seeded topics to capture them. Finally, we complement these $9$ keyword topics with $25$ additional unsupervised topics to accommodate the range of (potentially unanticipated) textual themes. 

\begin{figure}[t]
\centering
\includegraphics[width=1.\textwidth]{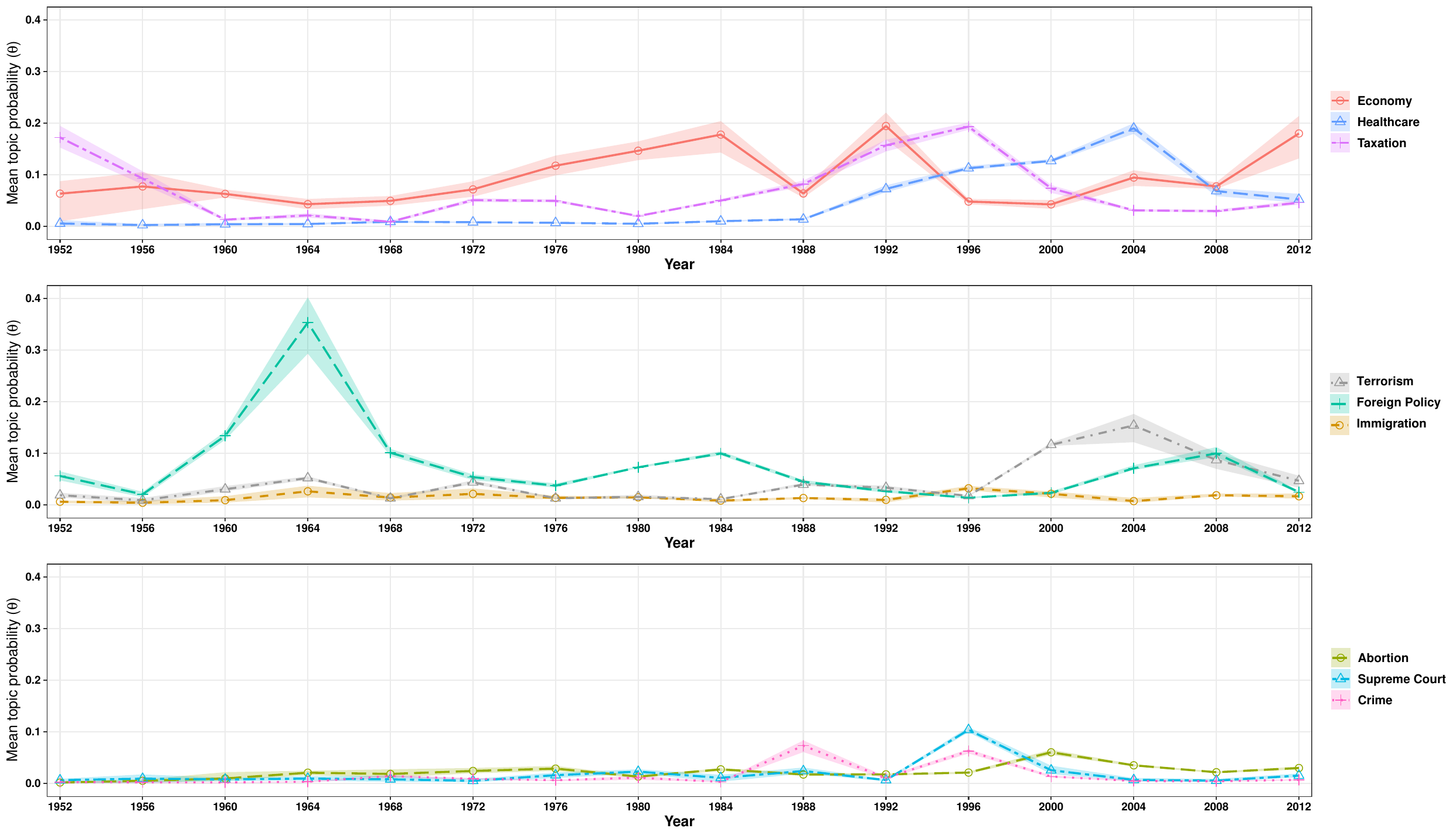} 
\caption[Issues over time]{\textbf{Evolution of key 2016-2024 election issues over time.} We plot topic prevalence (measured via mean topic probability) over presidential elections from $1952$-$2012$. Topics focus on $9$ key issue areas, which represent the union of the top $3$ issues from each of the most recent $3$ presidential elections, $2016$-$2024$, plus two additional topics (taxation and crime) that reflect remaining issues among top-$3$ issue areas in terms of advertiser attention. For presentation, we partition the issues into three plots corresponding to economic policy and social welfare, security and international affairs, and law and the courts. Error ribbons represent $90$\% credible intervals computed via the standard highest density interval method.}
\label{fig:topicmodel}
\end{figure}

Figure~\ref{fig:topicmodel} plots the estimated proportions of topics for the $9$ issue-area-focused topics over the seven decades of ads in the dataset. While an in-depth analysis is beyond the scope of this paper, we find that the estimated proportions of foreign policy issues are inversely correlated with domestic issues related to the economy and taxation.

For example, consistent with the theoretical conjecture that incumbents should emphasize economic issues during periods of prosperity \cite{vavreck2009message}, ads in the 1950s often focus on the booming postwar economy and tax policy. These domestic topics gave way to foreign policy issues during periods when insurgent candidates attempted to draw attention to Cold War developments or when incumbents highlighted their opponents’ foreign policy positions. The 1964 campaign is illustrative: a spike in the foreign policy topic reflects Lyndon B. Johnson’s relentless critique—exemplified by the iconic ``Daisy'' ad—of Barry Goldwater’s escalatory support for deploying limited tactical nuclear weapons against communists in the immediate wake of the Cuban Missile Crisis and Johnson’s recent endorsement of the Nuclear Test Ban Treaty. 

Economic and tax issues resurface during the stagflation era and again during Bill Clinton’s heresthetical \citep{riker1986art} pivot to the “it’s the economy, stupid” strategy. These topics recede once more as foreign policy dominated the 2000s, particularly as George W. Bush focused his 2004 reelection campaign overwhelmingly on the War on Terror, favoring bellicose imagery and interventionist military rhetoric against Kerry's more topically varied messaging that included a significant focus on healthcare reform (reflected in the spike in the healthcare topic in Fig. \ref{fig:topicmodel}, top plot). In 2008, Barack Obama and John McCain maintained the central focus on terrorism and foreign affairs, though by then the discourse had largely shifted toward troop withdrawal and accountability. Obama’s campaign concurrently increased focus on economic topics as he refocused the contest through his “Wall Street versus Main Street” refrain and attempted to catch challenger McCain flat-footed on issues related to the contemporaneous onset of the financial crisis.

We also observe the nascence, ebb, and flow of the now-pervasive `crime' issue area, beginning with the infamous `Horton weekend pass' dog whistle that aired during Bush's $1988$ campaign, and resurging in Bill Clinton's infamous `superpredator' and `three strikes' marketing. We note that political campaign supply-side measures of content such as these offer an interesting point of comparison to standard opinion polls that only capture the `demand-side' of political campaigns. 

\section*{Usage Notes}

In addition to bulk dataset access described in the \emph{Data Records} section above, we also provide a project website (\url{https://campaignTVads.org/}) with a convenient online interface designed to facilitate research by allowing practitioners to search and view videos that are relevant to their specific research questions. 

\subsection*{Videos by Candidate, Election, and/or Exact \& Fuzzy Transcript Keyword Search}

The ``Data'' tab of our project website provides drop-down menus that subset the videos by year and candidate's last name. It also provides a search bar that queries the transcripts of all the ads that were generated in \textbf{Step 1} above.  The search engine then returns videos whose transcripts contain either exact or fuzzy matches, with the former being returned first. The videos returned by the search engine are hosted on Vimeo, where they can be \textit{viewed online} and also individually \textit{downloaded} for further analysis. 
 
\section*{Code availability}

Our replication codes as well as codes for applying our workflow to other video datasets can be found on GitHub at: \url{https://github.com/AdamBreuer/AI-SummarizeVid}.

\section*{Acknowledgments} 

This project was funded by a grant from the National Science Foundation's Accountable Institutions and Behavior (AIB) program and the Established Program to Stimulate Competitive Research (EPSCoR). Award Numbers SES-2147635 and SES-2148928. Adam Breuer's research lab is funded by OpenAI. 

\section*{Author contributions statement}

J.P. led the archival team, and J.P. and M.C. directed the student coders at the University of Oklahoma. B.D. led the validation effort using student coders at the University of Iowa. A.B. created the preprocessing protocol, video summarization workflow, and the automated transcripts and summaries. M.B. created various online tools that helped in the coding effort, as well as the project website. A.B., J.P., and K.I. designed the statistical validation. A.B. conducted the empirical illustration with topic modeling. K.I. helped structure and develop all aspects of the project and manuscript. All authors reviewed the manuscript. 

\section*{Competing interests}
None of the authors have competing interests.

\appendix

\section{Appendix}

For completeness, we include the deferred Beta regression error model results from our validation section (eq., \ref{beta-preproc} and eq. \ref{beta-transcript}). These results are reported in Tables \ref{t:beta-preprocessing-reg} and \ref{t:beta-transcription-err}. Here, we show that ad preprocessing errors  and transcription errors have no statistically significant relationship with our stratification features, partisanship and year. 

\begin{table}[!htbp] \centering 
  \caption{Beta regression of preprocessing ad trimming errors (sec) on partisanship and year (eq. \ref{beta-preproc}).} 
  \label{t:beta-preprocessing-reg} 
\begin{tabular}{@{\extracolsep{5pt}}lccc} 
\\[-1.8ex]\hline 
\hline \\[-1.8ex] 
 & \multicolumn{3}{c}{\textit{Dependent variable:}} \\ 
\cline{2-4} 
\\[-1.8ex] & $\text{Error}_{start}$ & $\text{Error}_{end}$ & |$\text{Error}_{start}|+|\text{Error}_{end}|$ \\ 
\\[-1.8ex] & (1) & (2) & (3)\\ 
\hline \\[-1.8ex] 
 Republican & 1.955 & 4.260 & $-$46.664 \\ 
  & (24.059) & (38.836) & (44.025) \\ 
  & & & \\ 
 Election Year & 0.002 & 0.022 & $-$0.021 \\ 
  & (0.009) & (0.014) & (0.016) \\ 
  & & & \\ 
 Repub.*Elec.Year & $-$0.001 & $-$0.002 & 0.023 \\ 
  & (0.012) & (0.020) & (0.022) \\ 
  & & & \\ 
 Constant & $-$3.837 & $-$44.030 & 44.217 \\ 
  & (17.012) & (27.461) & (31.130) \\ 
  & & & \\ 
\hline \\[-1.8ex] 
Observations & 900 & 900 & 900 \\ 
R$^{2}$ & 0.002 & 0.006 & 0.004 \\ 
\hline 
\hline \\[-1.8ex] 
\textit{Note:}  & \multicolumn{3}{r}{$^{*}$p$<$0.1; $^{**}$p$<$0.05; $^{***}$p$<$0.01} \\ 
\end{tabular} 
\end{table} 
 
\begin{table}[!htbp] \centering 
  \caption{Beta regression of transcription word error fractions on partisanship and year  (eq. \ref{beta-transcript}).}  
  \label{t:beta-transcription-err} 
\begin{tabular}{@{\extracolsep{5pt}}lc} 
\\[-1.8ex]\hline 
\hline \\[-1.8ex] 
 & \multicolumn{1}{c}{\textit{Dependent variable:}} \\ 
\cline{2-2} 
\\[-1.8ex] & Transcription Error Fraction $R_j$ \\ 
\hline \\[-1.8ex] 
 Republican & 8.782 \\ 
  & (16.383) \\ 
  & \\ 
 Election Year & 0.001 \\ 
  & (0.006) \\ 
  & \\ 
 Repub.*Elec.Year & $-$0.004 \\ 
  & (0.008) \\ 
  & \\ 
 Constant & 1.565 \\ 
  & (11.588) \\ 
  & \\ 
\hline \\[-1.8ex] 
Observations & 119 \\ 
R$^{2}$ & 0.010 \\ 
\hline 
\hline \\[-1.8ex] 
\textit{Note:}  & \multicolumn{1}{r}{$^{*}$p$<$0.1; $^{**}$p$<$0.05; $^{***}$p$<$0.01} \\ 
\end{tabular} 
\end{table} 
 

\begin{thebibliography}{10}
\urlstyle{rm}
\expandafter\ifx\csname url\endcsname\relax
  \def\url#1{\texttt{#1}}\fi
\expandafter\ifx\csname urlprefix\endcsname\relax\def\urlprefix{URL }\fi
\expandafter\ifx\csname doiprefix\endcsname\relax\def\doiprefix{DOI: }\fi
\providecommand{\bibinfo}[2]{#2}
\providecommand{\eprint}[2][]{\url{#2}}

\bibitem{AdImpactPlay}
\bibinfo{title}{Adimpact: The play for the white house} (\bibinfo{year}{2024}).

\bibitem{AdImpactReview}
\bibinfo{title}{Adimpact: The 2024 cycle in review} (\bibinfo{year}{2024}).

\bibitem{borrell2017}
\bibinfo{author}{{Borrell Associates}}.
\newblock \bibinfo{title}{The final analysis: What happened to political
  advertising in 2016 (and forever)} (\bibinfo{year}{2017}).

\bibitem{jamieson1996packaging}
\bibinfo{author}{Jamieson, K.~H.}
\newblock \emph{\bibinfo{title}{Packaging the presidency: A history and
  criticism of presidential campaign advertising}} (\bibinfo{publisher}{Oxford
  University Press}, \bibinfo{year}{1996}).

\bibitem{kaid2004political}
\bibinfo{author}{Kaid, L.~L.}
\newblock \bibinfo{journal}{\bibinfo{title}{Political advertising}}.
\newblock {\emph{\JournalTitle{Handbook of political communication research}}}
  \textbf{\bibinfo{volume}{155202}} (\bibinfo{year}{2004}).

\bibitem{brader2006campaigning}
\bibinfo{author}{Brader, T.}
\newblock \emph{\bibinfo{title}{Campaigning for hearts and minds: How emotional
  appeals in political ads work}} (\bibinfo{publisher}{University of Chicago
  Press}, \bibinfo{year}{2006}).

\bibitem{biocca2013television}
\bibinfo{author}{Biocca, F.}
\newblock \emph{\bibinfo{title}{Television and political advertising: Volume I:
  Psychological processes}} (\bibinfo{publisher}{Routledge},
  \bibinfo{year}{2013}).

\bibitem{valentino2004impact}
\bibinfo{author}{Valentino, N.~A.}, \bibinfo{author}{Hutchings, V.~L.} \&
  \bibinfo{author}{Williams, D.}
\newblock \bibinfo{journal}{\bibinfo{title}{The impact of political advertising
  on knowledge, internet information seeking, and candidate preference}}.
\newblock {\emph{\JournalTitle{Journal of communication}}}
  \textbf{\bibinfo{volume}{54}}, \bibinfo{pages}{337--354}
  (\bibinfo{year}{2004}).

\bibitem{goldstein2002campaign}
\bibinfo{author}{Goldstein, K.} \& \bibinfo{author}{Freedman, P.}
\newblock \bibinfo{journal}{\bibinfo{title}{Campaign advertising and voter
  turnout: New evidence for a stimulation effect}}.
\newblock {\emph{\JournalTitle{Journal of Politics}}}
  \textbf{\bibinfo{volume}{64}}, \bibinfo{pages}{721--740}
  (\bibinfo{year}{2002}).

\bibitem{geer2008defense}
\bibinfo{author}{Geer, J.~G.}
\newblock \emph{\bibinfo{title}{In defense of negativity: Attack ads in
  presidential campaigns}} (\bibinfo{publisher}{University of Chicago Press},
  \bibinfo{year}{2008}).

\bibitem{gerber2011large}
\bibinfo{author}{Gerber, A.~S.}, \bibinfo{author}{Gimpel, J.~G.},
  \bibinfo{author}{Green, D.~P.} \& \bibinfo{author}{Shaw, D.~R.}
\newblock \bibinfo{journal}{\bibinfo{title}{How large and long-lasting are the
  persuasive effects of televised campaign ads? results from a randomized field
  experiment}}.
\newblock {\emph{\JournalTitle{American Political Science Review}}}
  \textbf{\bibinfo{volume}{105}}, \bibinfo{pages}{135--150}
  (\bibinfo{year}{2011}).

\bibitem{fowler2018political}
\bibinfo{author}{Fowler, E.~F.}, \bibinfo{author}{Franz, M.~M.} \&
  \bibinfo{author}{Ridout, T.~N.}
\newblock \emph{\bibinfo{title}{Political advertising in the United States}}
  (\bibinfo{publisher}{Routledge}, \bibinfo{year}{2018}).

\bibitem{petrocik2003issue}
\bibinfo{author}{Petrocik, J.~R.}, \bibinfo{author}{Benoit, W.~L.} \&
  \bibinfo{author}{Hansen, G.~J.}
\newblock \bibinfo{journal}{\bibinfo{title}{Issue ownership and presidential
  campaigning, 1952-2000}}.
\newblock {\emph{\JournalTitle{Political Science Quarterly}}}
  \textbf{\bibinfo{volume}{118}}, \bibinfo{pages}{599--626}
  (\bibinfo{year}{2003}).

\bibitem{hayes2005candidate}
\bibinfo{author}{Hayes, D.}
\newblock \bibinfo{journal}{\bibinfo{title}{Candidate qualities through a
  partisan lens: A theory of trait ownership}}.
\newblock {\emph{\JournalTitle{American Journal of Political Science}}}
  \textbf{\bibinfo{volume}{49}}, \bibinfo{pages}{908--923}
  (\bibinfo{year}{2005}).

\bibitem{fowler2016political}
\bibinfo{author}{Fowler, E.~F.}, \bibinfo{author}{Franz, M.~M.} \&
  \bibinfo{author}{Ridout, T.~N.}
\newblock \emph{\bibinfo{title}{Political Advertising in the United States}}
  (\bibinfo{publisher}{Westview Press}, \bibinfo{address}{Boulder, Colo.},
  \bibinfo{year}{2016}).

\bibitem{goldstein2002political}
\bibinfo{author}{Goldstein, K.}, \bibinfo{author}{Franz, M.} \&
  \bibinfo{author}{Ridout, T.}
\newblock \bibinfo{journal}{\bibinfo{title}{Political advertising in 2000}}.
\newblock {\emph{\JournalTitle{Combined File [dataset]. Final release. Madison,
  WI: The Department of Political Science at The University of
  Wisconsin-Madison and the The Brennan Center for Justice at New York
  University}}}  (\bibinfo{year}{2002}).

\bibitem{goldstein2007congressional}
\bibinfo{author}{Goldstein, K.} \& \bibinfo{author}{Rivlin, J.}
\newblock \bibinfo{journal}{\bibinfo{title}{Congressional and gubernatorial
  advertising, 2003-2004}}.
\newblock {\emph{\JournalTitle{Combined File [dataset]. Final release. Madison,
  WI: The University of Wisconsin Advertising Project, The Department of
  Political Science at The University of Wisconsin-Madison}}}
  (\bibinfo{year}{2007}).

\bibitem{prior2001weighted}
\bibinfo{author}{Prior, M.}
\newblock \bibinfo{journal}{\bibinfo{title}{Weighted content analysis of
  political advertisements}}.
\newblock {\emph{\JournalTitle{Political Communication}}}
  \textbf{\bibinfo{volume}{18}}, \bibinfo{pages}{335--345}
  (\bibinfo{year}{2001}).

\bibitem{kanterpcc}
\bibinfo{title}{Carl albert center julian p. kanter political commercial
  collection}.
\newblock \bibinfo{howpublished}{\url{https://arc.ou.edu/}}.
\newblock \bibinfo{note}{Accessed on September 26, 2024}.

\bibitem{tarr:hwan:imai:23}
\bibinfo{author}{Tarr, A.}, \bibinfo{author}{Hwang, J.} \&
  \bibinfo{author}{Imai, K.}
\newblock \bibinfo{journal}{\bibinfo{title}{Automated coding of political
  campaign advertisement videos: An empirical validation study}}.
\newblock {\emph{\JournalTitle{Political Analysis}}}
  \textbf{\bibinfo{volume}{31}}, \bibinfo{pages}{554–574},
  \url{10.1017/pan.2022.26} (\bibinfo{year}{2023}).

\bibitem{harp201660}
\bibinfo{author}{Harp, A.~B.}
\newblock \bibinfo{journal}{\bibinfo{title}{The 60 second candidate}}.
\newblock {\emph{\JournalTitle{Sooner Magazine}}}
  \textbf{\bibinfo{volume}{36}}, \bibinfo{pages}{8--11} (\bibinfo{year}{2016}).

\bibitem{haynes1996political}
\bibinfo{author}{Haynes, K.}, \bibinfo{author}{Kaid, L.} \&
  \bibinfo{author}{Rand, C.}
\newblock \bibinfo{journal}{\bibinfo{title}{The political commercial archive:
  Management of moving image and sound recordings}}.
\newblock {\emph{\JournalTitle{The American Archivist}}}
  \textbf{\bibinfo{volume}{59}}, \bibinfo{pages}{48--61}
  (\bibinfo{year}{1996}).

\bibitem{pryse2022practical}
\bibinfo{author}{Pryse, J.}
\newblock \bibinfo{journal}{\bibinfo{title}{Practical remote workflow solutions
  for complex digital projects: Opportunities in a pandemic}}.
\newblock {\emph{\JournalTitle{Collections}}} \textbf{\bibinfo{volume}{18}},
  \bibinfo{pages}{258--279} (\bibinfo{year}{2022}).

\bibitem{clinton1992living}
\bibinfo{author}{Clinton, W.~J.}
\newblock \bibinfo{journal}{\bibinfo{title}{The living room candidate:
  Presidential campaign commercials: 1952--2004}}.
\newblock {\emph{\JournalTitle{American Museum of the Moving Image}}}
  (\bibinfo{year}{1992}).

\bibitem{kaid1991negative}
\bibinfo{author}{Kaid, L.~L.} \& \bibinfo{author}{Johnston, A.}
\newblock \bibinfo{journal}{\bibinfo{title}{Negative versus positive television
  advertising in us presidential campaigns, 1960-1988.}}
\newblock {\emph{\JournalTitle{Journal of communication}}}
  \textbf{\bibinfo{volume}{41}}, \bibinfo{pages}{53--64}
  (\bibinfo{year}{1991}).

\bibitem{finkel1998spot}
\bibinfo{author}{Finkel, S.~E.} \& \bibinfo{author}{Geer, J.~G.}
\newblock \bibinfo{journal}{\bibinfo{title}{A spot check: Casting doubt on the
  demobilizing effect of attack advertising}}.
\newblock {\emph{\JournalTitle{American journal of political science}}}
  \bibinfo{pages}{573--595} (\bibinfo{year}{1998}).

\bibitem{damore2003using}
\bibinfo{author}{Damore, D.~F.}
\newblock \bibinfo{journal}{\bibinfo{title}{Using campaign advertisements to
  assess campaign processes}}.
\newblock {\emph{\JournalTitle{Journal of Political Marketing}}}
  \textbf{\bibinfo{volume}{3}}, \bibinfo{pages}{39--59} (\bibinfo{year}{2003}).

\bibitem{prior2007post}
\bibinfo{author}{Prior, M.}
\newblock \emph{\bibinfo{title}{Post-broadcast democracy: How media choice
  increases inequality in political involvement and polarizes elections}}
  (\bibinfo{publisher}{Cambridge University Press}, \bibinfo{year}{2007}).

\bibitem{brader2020campaigning}
\bibinfo{author}{Brader, T.}
\newblock \emph{\bibinfo{title}{Campaigning for hearts and minds: How emotional
  appeals in political ads work}} (\bibinfo{publisher}{University of Chicago
  Press}, \bibinfo{year}{2020}).

\bibitem{schofield2016amicus}
\bibinfo{author}{Schofield, B.~L.} \& \bibinfo{author}{Bailey, L.}
\newblock \bibinfo{title}{Amicus brief in support of
  defendant-appellant-cross-appellee in fox news v. tv eyes}
  (\bibinfo{year}{2016}).

\bibitem{power}
\bibinfo{author}{Colt, S.~{\relax (Director) Dobrowksi, H. {\relax (Producer)}
  }.}
\newblock \bibinfo{title}{Prayer. politics. power}.
\newblock \bibinfo{howpublished}{A Sarah Colt Productions Film for American
  Experience} (\bibinfo{year}{2015}).

\bibitem{octopus}
\bibinfo{author}{Treitz, Z.~{\relax (Director) Ambrose, M. {\relax (Producer)}
  }.}
\newblock \bibinfo{title}{American conspiracy: The octopus murders}
  (\bibinfo{year}{2024}).

\bibitem{wilder}
\bibinfo{author}{MacNicol, G.}
\newblock \bibinfo{title}{Wilder: Daughter dearest pt. 2: Politics and rose}.
\newblock \bibinfo{howpublished}{I Heart Radio} (\bibinfo{year}{June 29,
  2023}).

\bibitem{beard2020advertising}
\bibinfo{author}{Beard, F.} \& \bibinfo{author}{Timke, E.}
\newblock \bibinfo{journal}{\bibinfo{title}{Advertising in the archives: A
  guide to advertising and marketing archives around the world}}.
\newblock {\emph{\JournalTitle{Advertising \& Society Quarterly}}}
  \textbf{\bibinfo{volume}{21}} (\bibinfo{year}{2020}).

\bibitem{dover2002videostyle}
\bibinfo{author}{Dover, E.}
\newblock \bibinfo{journal}{\bibinfo{title}{Videostyle in presidential
  campaigns: Style and content of televised political advertising}}.
\newblock {\emph{\JournalTitle{Great Plains Quarterly}}}
  \textbf{\bibinfo{volume}{22}}, \bibinfo{pages}{60} (\bibinfo{year}{2002}).

\bibitem{bernard2020archival}
\bibinfo{author}{Bernard, S.~C.} \& \bibinfo{author}{Rabin, K.}
\newblock \emph{\bibinfo{title}{Archival storytelling: a filmmaker’s guide to
  finding, using, and licensing third-party visuals and music}}
  (\bibinfo{publisher}{Routledge}, \bibinfo{year}{2020}).

\bibitem{biocca1991television}
\bibinfo{author}{Biocca, F.} \emph{et~al.}
\newblock \bibinfo{journal}{\bibinfo{title}{Television and political
  advertising}}.
\newblock {\emph{\JournalTitle{Taylor \& Francis eBooks DRM Free Collection}}}
  (\bibinfo{year}{1991}).

\bibitem{dalton2011third}
\bibinfo{author}{Dalton, P.} \& \bibinfo{author}{McIlwain, C.}
\newblock \bibinfo{journal}{\bibinfo{title}{Third-party “hatchet” ads: An
  exploratory content study comparing third-party and candidate spots from the
  2004 presidential election}}.
\newblock {\emph{\JournalTitle{Atlantic Journal of Communication}}}
  \textbf{\bibinfo{volume}{19}}, \bibinfo{pages}{129--151}
  (\bibinfo{year}{2011}).

\bibitem{biocca2014television}
\bibinfo{author}{Biocca, F.}
\newblock \emph{\bibinfo{title}{Television and Political Advertising: Volume
  II: Signs, Codes, and Images}} (\bibinfo{publisher}{Routledge},
  \bibinfo{year}{2014}).

\bibitem{dunaway2019effects}
\bibinfo{author}{Dunaway, J.}, \bibinfo{author}{Searles, K.},
  \bibinfo{author}{Fowler, E.}, \bibinfo{author}{Ridout, T.} \&
  \bibinfo{author}{Napoli, P.}
\newblock \bibinfo{journal}{\bibinfo{title}{The effects of political
  advertising: assessing the impact of changing technologies, strategies and
  tactics}}.
\newblock {\emph{\JournalTitle{Mediated communication: Handbooks of
  communication science}}} \textbf{\bibinfo{volume}{7}}, \bibinfo{pages}{1--22}
  (\bibinfo{year}{2019}).

\bibitem{webb2008freedom}
\bibinfo{author}{Webb, C.}
\newblock \bibinfo{journal}{\bibinfo{title}{Freedom for all? blacks, jews, and
  the political censorship of white racists in the civil rights era}}.
\newblock {\emph{\JournalTitle{American Jewish History}}}
  \textbf{\bibinfo{volume}{94}}, \bibinfo{pages}{267--297}
  (\bibinfo{year}{2008}).

\bibitem{nelson1989sources}
\bibinfo{author}{Nelson, R.~A.}
\newblock \bibinfo{journal}{\bibinfo{title}{Sources for archival research on
  film and television propaganda in the united states}}.
\newblock {\emph{\JournalTitle{Film History}}} \bibinfo{pages}{333--340}
  (\bibinfo{year}{1989}).

\bibitem{albitz2001locating}
\bibinfo{author}{Albitz, R.~S.}
\newblock \bibinfo{journal}{\bibinfo{title}{Locating moving image materials for
  multimedia development: A reference strategy}}.
\newblock {\emph{\JournalTitle{The Reference Librarian}}}
  \textbf{\bibinfo{volume}{34}}, \bibinfo{pages}{99--110}
  (\bibinfo{year}{2001}).

\bibitem{grimmer2022text}
\bibinfo{author}{Grimmer, J.}, \bibinfo{author}{Roberts, M.~E.} \&
  \bibinfo{author}{Stewart, B.~M.}
\newblock \emph{\bibinfo{title}{Text as data: A new framework for machine
  learning and the social sciences}} (\bibinfo{publisher}{Princeton University
  Press}, \bibinfo{year}{2022}).

\bibitem{eshima2024keyword}
\bibinfo{author}{Eshima, S.}, \bibinfo{author}{Imai, K.} \&
  \bibinfo{author}{Sasaki, T.}
\newblock \bibinfo{journal}{\bibinfo{title}{Keyword-assisted topic models}}.
\newblock {\emph{\JournalTitle{American Journal of Political Science}}}
  \textbf{\bibinfo{volume}{68}}, \bibinfo{pages}{730--750}
  (\bibinfo{year}{2024}).

\bibitem{egami2022make}
\bibinfo{author}{Egami, N.}, \bibinfo{author}{Fong, C.~J.},
  \bibinfo{author}{Grimmer, J.}, \bibinfo{author}{Roberts, M.~E.} \&
  \bibinfo{author}{Stewart, B.~M.}
\newblock \bibinfo{journal}{\bibinfo{title}{How to make causal inferences using
  texts}}.
\newblock {\emph{\JournalTitle{Science Advances}}}
  \textbf{\bibinfo{volume}{8}}, \bibinfo{pages}{eabg2652}
  (\bibinfo{year}{2022}).

\bibitem{fong2023causal}
\bibinfo{author}{Fong, C.} \& \bibinfo{author}{Grimmer, J.}
\newblock \bibinfo{journal}{\bibinfo{title}{Causal inference with latent
  treatments}}.
\newblock {\emph{\JournalTitle{American Journal of Political Science}}}
  \textbf{\bibinfo{volume}{67}}, \bibinfo{pages}{374--389}
  (\bibinfo{year}{2023}).

\bibitem{freedman1999measuring}
\bibinfo{author}{Freedman, P.} \& \bibinfo{author}{Goldstein, K.}
\newblock \bibinfo{journal}{\bibinfo{title}{Measuring media exposure and the
  effects of negative campaign ads}}.
\newblock {\emph{\JournalTitle{American journal of political Science}}}
  \bibinfo{pages}{1189--1208} (\bibinfo{year}{1999}).

\bibitem{goldsteinstoryboardbook}
\bibinfo{author}{Goldstein, K.~M.}, \bibinfo{author}{Freedman, P.~B.} \&
  \bibinfo{author}{Franz, M.~M.}
\newblock \emph{\bibinfo{title}{Campaign Advertising and American Democracy}}
  (\bibinfo{publisher}{Temple University Press},
  \bibinfo{address}{Philadelphia, PA}, \bibinfo{year}{2009}).

\bibitem{googleAPI}
\bibinfo{author}{Li, F.-F.}
\newblock \bibinfo{journal}{\bibinfo{title}{Announcing google cloud video
  intelligence api, and more cloud machine learning updates}}.
\newblock {\emph{\JournalTitle{arXiv preprint arXiv:1102.3975}}}
  (\bibinfo{year}{2017}).

\bibitem{radford2023robust}
\bibinfo{author}{Radford, A.} \emph{et~al.}
\newblock \bibinfo{title}{Robust speech recognition via large-scale weak
  supervision}.
\newblock In \emph{\bibinfo{booktitle}{International conference on machine
  learning}}, \bibinfo{pages}{28492--28518} (\bibinfo{organization}{PMLR},
  \bibinfo{year}{2023}).

\bibitem{fowler2021politicalonlineoffline}
\bibinfo{author}{Fowler, E.~F.}, \bibinfo{author}{Franz, M.~M.},
  \bibinfo{author}{Martin, G.~J.}, \bibinfo{author}{Peskowitz, Z.} \&
  \bibinfo{author}{Ridout, T.~N.}
\newblock \bibinfo{journal}{\bibinfo{title}{Political advertising online and
  offline}}.
\newblock {\emph{\JournalTitle{American Political Science Review}}}
  \textbf{\bibinfo{volume}{115}}, \bibinfo{pages}{130--149}
  (\bibinfo{year}{2021}).

\bibitem{walker1996mpi}
\bibinfo{author}{Walker, D.~W.} \& \bibinfo{author}{Dongarra, J.~J.}
\newblock \bibinfo{journal}{\bibinfo{title}{Mpi: a standard message passing
  interface}}.
\newblock {\emph{\JournalTitle{Supercomputer}}} \textbf{\bibinfo{volume}{12}},
  \bibinfo{pages}{56--68} (\bibinfo{year}{1996}).

\bibitem{gabriel2004open}
\bibinfo{author}{Gabriel, E.} \emph{et~al.}
\newblock \bibinfo{title}{Open mpi: Goals, concept, and design of a next
  generation mpi implementation}.
\newblock In \emph{\bibinfo{booktitle}{Recent Advances in Parallel Virtual
  Machine and Message Passing Interface: 11th European PVM/MPI Users’ Group
  Meeting Budapest, Hungary, September 19-22, 2004. Proceedings 11}},
  \bibinfo{pages}{97--104} (\bibinfo{organization}{Springer},
  \bibinfo{year}{2004}).

\bibitem{dalcin2021mpi4py}
\bibinfo{author}{Dalcin, L.} \& \bibinfo{author}{Fang, Y.-L.~L.}
\newblock \bibinfo{journal}{\bibinfo{title}{mpi4py: Status update after 12
  years of development}}.
\newblock {\emph{\JournalTitle{Computing in Science \& Engineering}}}
  \textbf{\bibinfo{volume}{23}}, \bibinfo{pages}{47--54}
  (\bibinfo{year}{2021}).

\bibitem{ctap2025}
\bibinfo{author}{Breuer, A.}, \bibinfo{author}{Butler, M.},
  \bibinfo{author}{Crespin, M.~H.}, \bibinfo{author}{Dietrich, B.~J.} \&
  \bibinfo{author}{Imai, K.}
\newblock \bibinfo{journal}{\bibinfo{title}{Campaign television advertisement
  project (\mbox{CTAP}) dataset}}.
\newblock \url{https://doi.org/10.7910/DVN/7CJSLD} (\bibinfo{year}{2025}).
\newblock \bibinfo{note}{DOI is associated with the first year of data
  available. Full archive \& all years available at
  \url{https://dataverse.harvard.edu/dataverse/ctap}}.

\bibitem{mcgraw1996forming}
\bibinfo{author}{McGraw, K.~O.} \& \bibinfo{author}{Wong, S.~P.}
\newblock \bibinfo{journal}{\bibinfo{title}{Forming inferences about some
  intraclass correlation coefficients.}}
\newblock {\emph{\JournalTitle{Psychological methods}}}
  \textbf{\bibinfo{volume}{1}}, \bibinfo{pages}{30} (\bibinfo{year}{1996}).

\bibitem{hamed2023benchmarking}
\bibinfo{author}{Hamed, I.} \emph{et~al.}
\newblock \bibinfo{title}{Benchmarking evaluation metrics for code-switching
  automatic speech recognition}.
\newblock In \emph{\bibinfo{booktitle}{2022 IEEE Spoken Language Technology
  Workshop (SLT)}}, \bibinfo{pages}{999--1005} (\bibinfo{organization}{IEEE},
  \bibinfo{year}{2023}).

\bibitem{smithson2006better}
\bibinfo{author}{Smithson, M.} \& \bibinfo{author}{Verkuilen, J.}
\newblock \bibinfo{journal}{\bibinfo{title}{A better lemon squeezer?
  maximum-likelihood regression with beta-distributed dependent variables.}}
\newblock {\emph{\JournalTitle{Psychological methods}}}
  \textbf{\bibinfo{volume}{11}}, \bibinfo{pages}{54} (\bibinfo{year}{2006}).

\bibitem{smithson2006bettersupp}
\bibinfo{author}{Smithson, M.} \& \bibinfo{author}{Verkuilen, J.}
\newblock \bibinfo{journal}{\bibinfo{title}{Supplementary materials for: A
  better lemon squeezer? maximum-likelihood regression with beta-distributed
  dependent variables.}}
\newblock {\emph{\JournalTitle{Psychological methods}}}
  \textbf{\bibinfo{volume}{11}}, \bibinfo{pages}{54} (\bibinfo{year}{2006}).

\bibitem{kryscinski2019neural}
\bibinfo{author}{Kry{\'s}ci{\'n}ski, W.}, \bibinfo{author}{Keskar, N.~S.},
  \bibinfo{author}{McCann, B.}, \bibinfo{author}{Xiong, C.} \&
  \bibinfo{author}{Socher, R.}
\newblock \bibinfo{title}{Neural text summarization: A critical evaluation}.
\newblock In \emph{\bibinfo{booktitle}{Proceedings of the 2019 Conference on
  Empirical Methods in Natural Language Processing and the 9th International
  Joint Conference on Natural Language Processing (EMNLP-IJCNLP)}},
  \bibinfo{pages}{540--551}, \url{10.18653/v1/D19-1051}
  (\bibinfo{publisher}{Association for Computational Linguistics},
  \bibinfo{address}{Hong Kong, China}, \bibinfo{year}{2019}).

\bibitem{fabbri2021summeval}
\bibinfo{author}{Fabbri, A.~R.} \emph{et~al.}
\newblock \bibinfo{journal}{\bibinfo{title}{Summeval: Re-evaluating
  summarization evaluation}}.
\newblock {\emph{\JournalTitle{Transactions of the Association for
  Computational Linguistics}}} \textbf{\bibinfo{volume}{9}},
  \bibinfo{pages}{391--409} (\bibinfo{year}{2021}).

\bibitem{vavreck2009message}
\bibinfo{author}{Vavreck, L.}
\newblock \emph{\bibinfo{title}{The Message Matters: The Economy and
  Presidential Campaigns}}.
\newblock Political science : American politics : Economics
  (\bibinfo{publisher}{Princeton University Press}, \bibinfo{year}{2009}).

\bibitem{pew2024}
\bibinfo{author}{Doherty, C.}, \bibinfo{author}{Kiley, J.} \&
  \bibinfo{author}{Asheer, N.}
\newblock \bibinfo{journal}{\bibinfo{title}{In tied presidential race, harris
  and trump have contrasting strengths, weaknesses}}.
\newblock {\emph{\JournalTitle{Pew Research Center}}}
  \textbf{\bibinfo{volume}{September}} (\bibinfo{year}{2024}).

\bibitem{pew2020}
\bibinfo{author}{Doherty, C.}, \bibinfo{author}{Kiley, J.},
  \bibinfo{author}{Asheer, N.} \& \bibinfo{author}{Jordan, C.}
\newblock \bibinfo{journal}{\bibinfo{title}{Election 2020: Voters are highly
  engaged, but nearly half expect to have difficulties voting}}.
\newblock {\emph{\JournalTitle{Pew Research Center}}}
  \textbf{\bibinfo{volume}{August}} (\bibinfo{year}{2020}).

\bibitem{pew2016}
\bibinfo{author}{Doherty, C.}, \bibinfo{author}{Kiley, J.} \&
  \bibinfo{author}{Johnson, B.}
\newblock \bibinfo{journal}{\bibinfo{title}{2016 campaign: Strong interest,
  widespread dissatisfaction}}.
\newblock {\emph{\JournalTitle{Pew Research Center}}}
  \textbf{\bibinfo{volume}{July}} (\bibinfo{year}{2016}).

\bibitem{riker1986art}
\bibinfo{author}{Riker, W.}, \bibinfo{author}{Riker, W.} \&
  \bibinfo{author}{Riker, W.}
\newblock \emph{\bibinfo{title}{The Art of Political Manipulation}}.
\newblock Political science (\bibinfo{publisher}{Yale University Press},
  \bibinfo{year}{1986}).

\end{thebibliography}
\end{document}